\documentclass[sigconf]{acmart}

\AtBeginDocument{%
  \providecommand\BibTeX{{%
    \normalfont B\kern-0.5em{\scshape i\kern-0.25em b}\kern-0.8em\TeX}}}

\acmYear{2023}\copyrightyear{2023}
\setcopyright{acmlicensed}
\acmConference[ICS '23]{International Conference on Supercomputing}{June 21--23, 2023}{Orlando, FL, USA}
\acmBooktitle{International Conference on Supercomputing (ICS '23), June 21--23, 2023, Orlando, FL, USA}
\acmPrice{15.00}
\acmDOI{10.1145/3577193.3593725}
\acmISBN{979-8-4007-0056-9/23/06}

\usepackage[normalem]{ulem}

\usepackage{amsmath,amsfonts} 

\usepackage{caption}
\usepackage{subcaption}
\usepackage{graphicx}
\usepackage{textcomp}
\usepackage{xcolor}
\usepackage{multirow}
\usepackage{booktabs}

\usepackage{tikz}
\newcommand*\circled[1]{\tikz[baseline=(myanchor.base)]\node[circle,fill=.,inner sep=1pt](myanchor){\color{-.}\bfseries\footnotesize #1};}

\usepackage[linesnumbered,ruled,vlined]{algorithm2e}
\SetAlFnt{\small}
\newcommand\codefont[1]{\textcolor{red}{\texttt{\textbf{#1}}}} 

\SetCommentSty{mycommfont}
\SetKwInput{KwInput}{Input}                
\SetKwInput{KwOutput}{Output}              

\begin{document}

\title{BitGNN: Unleashing the Performance Potential of Binary Graph Neural Networks on GPUs}

\author{Jou-An Chen}
\affiliation{%
  \institution{North Carolina State University}
  \city{Raleigh}
  \country{NC, USA}}
\email{jchen73@ncsu.edu}

\author{Hsin-Hsuan Sung}
\affiliation{%
  \institution{North Carolina State University}
  \city{Raleigh}
  \country{NC, USA}}
\email{hsung2@ncsu.edu}

\author{Xipeng Shen}
\affiliation{%
  \institution{North Carolina State University}
  \city{Raleigh}
  \country{NC, USA}}
\email{xshen5@ncsu.edu}

\author{Sutanay Choudhury}
\affiliation{%
  \institution{Pacific Northwest National Laboratory}
  \city{Richland}
  \country{WA, USA}}
\email{Sutanay.Choudhury@pnnl.gov}

\author{Ang Li}
\affiliation{%
  \institution{Pacific Northwest National Laboratory}
  \city{Richland}
  \country{WA, USA}}
\email{ang.li@pnnl.gov}

\begin{abstract}
Recent studies have shown that Binary Graph Neural Networks (GNNs) are promising for saving computations of GNNs through binarized tensors. Prior work, however, mainly focused on algorithm designs or training techniques, leaving it open to how to materialize the performance potential on accelerator hardware fully. This work redesigns the binary GNN inference backend from the efficiency perspective. It fills the gap by proposing a series of abstractions and techniques to map binary GNNs and their computations best to fit the nature of bit manipulations on GPUs. Results on real-world graphs with GCNs, GraphSAGE, and GraphSAINT show that the proposed techniques outperform state-of-the-art binary GNN implementations by 8-22X with the same accuracy maintained. BitGNN code is publicly available.\footnote{https://github.com/PICTureRG/BitGNN}.
\end{abstract}



\maketitle
\section{Introduction}
Recent years have witnessed a rapidly increasing adoption of Graph Neural Networks (GNNs) in various domains, from social networks~\cite{fan2019graph} to bioinformatics~\cite{reau2023deeprank}, computational chemistry~\cite{gilmer2017neural, helal2022extreme}, 3D computer vision~\cite{wang2019dynamic, zhang2021pointx}, and so on. GNN allows the linear and non-linear transformations of Neural Networks to work directly on graphs. By seamlessly integrating both the graph structure and node/edge features into the modeling, GNN is a natural fit for graph-based problems. Numerous studies~\cite{kipf2017semi, yao2019graph, ye2019web, liang2020deep, wan2020hyperspectral, jiang2020hi} on graph-based classification and prediction have reported significantly better results achieved through GNNs than traditional methods. 

Because real-world graphs are often large, GNN inference is often time-consuming and space-hungry, frequently exceeding the capacity of the storage or memory and the desirable latency. Inspired by the binarization (i.e., quantization to the 1-bit extreme) for ordinary deep neural networks~\cite{hubara2016binarized, courbariaux2016binarized, rastegari2016xnor, bulat2019xnor, zhang2022pokebnn}, an emerging effort is to investigate the potential of GNN binarization. By converting each value in the activation maps, weights, and/or adjacency matrices into a single bit (sometimes through matrix factorizations), GNN binarization can significantly reduce the space demands and computation amounts. 

The continuous research in the recent several years on GNN binarization has achieved some remarkable progress~\cite{wang2021bi, bahri2021binary, wang2021binarized, jing2021meta}. For instance, the accuracy loss caused by binarization has reduced from 16\% to less than 5\%~\cite{wang2021bi}. These studies show solutions that can control the loss of accuracy within a small percentage on various graphs and architectures while giving significant theoretical reductions in the number of computations and memory usage. Nevertheless, how to turn the potential into full speedups and memory savings on real machines remains an open question. For example, based on a theoretical analysis, the authors of a recent study~\cite{wang2021bi} on binary GNNs report $\sim$47X potential savings of computations and $\sim$30X possible memory footprint reduction. Still, their experiments show no speedups or memory savings over the original non-binarized GNN.

Existing work on binary GNNs mainly focuses on {\em algorithm-level} improvement on network architectures~\cite{wang2021bi, bahri2021binary, wang2021binarized} or training techniques~\cite{wang2021bi, wang2021binarized, jing2021meta}. The immense potential of binary GNN inference on hardware has not yet been  harvested.  

Quantizing tensors into bits and representing and manipulating them efficiently on word-based architecture involves lots of intricacies. It is even more so for GNNs, for several reasons. First, a GNN layer typically consists of interleaving dense and sparse operators. It is not straightforward to reconcile the inconsistency between dense and sparse bit-level layouts while minimizing the inference latency and achieving desirable accuracy-speed tradeoffs. Second, in binary GNNs, the meaning of bits in activations/weights (1/0 means +/-) and graphs (1/0 means connectivity) differ. For these special properties, integer intrinsics should be carefully orchestrated to maximize bit-level and thread-level parallelism. Third, to fully harness the significant memory reduction brought by bit representation, it is important to synergize the careful use of registers and memory hierarchy with  the massive parallelism on accelerators.

This work provides the first known systematic exploration to unlock the performance potential of binary GNNs. The target hardware it focuses on is Graphics Processing Units (GPUs), the most commonly used device for GNNs and binary GNNs. This work makes four-fold contributions: 

\noindent\textbf{(1) Materializing binary GNNs with a series of bit-level optimized abstraction.} To effectively materialize a binary GNN model, there are various possible compositions of bit-level BLAS kernels with various choices in precisions, eliminations of rebinarization, and fusions. We propose the first known two-level Binary GNN abstraction, which hides the complexities from users, offers flexible support of various scenarios and enables easy drop-in replacements for converting a GNN into a binary GNN.  

\noindent\textbf{(2) Representation of bit tensors.} Binary GNN includes binarized activations, weights, and adjacency matrices. Unlike activations and weights, adjacency matrices are typically sparse, the representation of which is subject to a granularity dilemma. We propose a {\em fine-representing dynamic-coarsening} (FRDC) scheme to simultaneously minimize graph storage cost and maximize kernel efficiency on a GPU. Additionally, we propose several solutions to reconcile the 1/0 and +/- inconsistency between binary graphs and activations.

\noindent\textbf{(3) Materialization of the BitGNN abstractions.} We develop the BitGNN abstractions into an efficient library through careful bit manipulations via binary intrinsics and memory and parallelism optimizations. We devise and materialize the abstractions for different precision needs with bit-manipulation intrinsics. We propose a series of techniques with dedicated treatment on workload decomposition, bit-tensor load/store with registers and on-chip memory usage, intra-warp and inter-warp synchronizations, and efficient binary sparse matrix multiplication (BSpMM) kernels.

\noindent\textbf{(4) Evaluation of BitGNN.} We integrate all the techniques into a full solution, namely BitGNN, in the form of a collection of a library and tuning utilities. Compared to the state-of-the-art binary GNN~\cite{wang2021bi} and full-precision GNN~\cite{fey2019fast} implementations, BitGNN gives up to 26X, 22X, and 19X latency reduction on GCN, GraphSAGE and GraphSAINT models. Meanwhile, BitGNN makes GNNs possible to efficiently process several graphs that are too large to process by prior solutions. 
\section{Background}

\subsection{Graph Neural Networks}
There are typically two types of operations in a GNN: \textit{aggregate} and \textit{apply}. The \textit{aggregate} phase collects the information from the input graph's geometric structure. The \textit{apply} phase performs linear or non-linear operations as in typical Neural Networks. A GNN can be formulated as follows:

\[\mathbf{X}^{(l+1)}_i = \gamma_{\mathbf{\theta}} \left( \mathbf{X}^{(l)}_i,
\Psi_{j \in \mathcal{N}(i)} \, \phi_{\mathbf{\theta}}
\left(\mathbf{X}^{(l)}_i, \mathbf{X}^{(l)}_j,\mathbf{E}_{i,j}\right) \right)\]

For node classification tasks, the node representation is the activation in each GNN layer. The representation of node $i$ at layer $l$ is denoted as $\mathbf{X}^{(l)}_i$. $\mathcal{N}(i)$ represents the set of neighbors of node $i$. In an \textit{aggregate} phase, $\Psi$ denotes a permutation-invariant function that performs message aggregation. The commonly used functions include sum, mean, and max. Before the aggregation, $\phi_{\mathbf{\theta}}$ denotes any user-defined operators applied to the following: $\mathbf{X}_i$ (the representation of node $i$), $\mathbf{X}_j$ (the representation of node $j$, the neighbors of node $i$), or $\mathbf{E}_{i,j}$ (the representation of edge $(i, j)$). After the aggregation, $\gamma_{\mathbf{\theta}}$ denotes the operations of the \textit{apply} phase. Most GNNs have neural operations (e.g., MLP and ReLU) in this phase. We briefly explain several popular GNNs.

\textbf{GCNConv} Graph Convolutional Network (GCN)~\cite{kipf2017semi} introduces a semi-supervised learning algorithm that can learn the graph structure directly. A forward layer of GCNConv can be defined as follows:

\[\mathbf{X}^{(l+1)} = \sigma(\tilde{A}\mathbf{X}^{(l)}\mathbf{W}^{(l)}),\]

\noindent where, the normalized adjacency matrix $\tilde{A}$ equals $\hat{D}^{-\frac{1}{2}}\hat{A}\hat{D}^{-\frac{1}{2}}$. The $\hat{A} = A + I$ is the adjacency matrix plus self-adjacency, and $\hat{D}_{ii} = \sum_{j=0} \hat{A}_{ij}$ is its diagonal out-degree matrix. $X^{(l)}\in\mathbb{R}^{N \times d}$ denotes the node feature embedding in the $l$-th layer, and $W^{(l)}$ is the learnable parameters or weights, and $\sigma$ represents a non-linear activation function (e.g., ReLU). Bi-GCN~\cite{wang2021bi} is composed of two such GCNConv layers.

\textbf{SAGEConv} SAGEConv~\cite{hamilton2017inductive} enhances the graph convolution layer with a mean aggregator as it is a linear approximation of a localized spectral convolution~\cite{kipf2017semi}. It also extends the layer with a skip connection to emulate the concatenation of the prior layer's node representation in the convolutional aggregator. A SAGEConv layer is defined as follows:

\[\mathbf{X}_i^{(l+1)} = \mathbf{X}_i^{(l)} \mathbf{W}^{(l)}_1 + \mathrm{mean}_{j \in \mathcal{N}(i)} \mathbf{X}^{(l)}_j \cdot \mathbf{W}^{(l)}_2\]

\textbf{GraphConv} GraphConv~\cite{morris2019weisfeiler} is similar to SAGEConv except that it uses a normal sum aggregator:

\[\mathbf{X}_i^{(l+1)} = \mathbf{X}_i^{(l)} \mathbf{W}^{(l)}_1 + \sum_{j \in \mathcal{N}(i)} \mathbf{X}_j^{(l)} \mathbf{W}^{(l)}_2\]

To allow learning on large graphs, GraphSAGE~\cite{hamilton2017inductive} proposes a neighbor sampling approach to deal with the neighbor explosion problem on large graph training. Bi-GraphSAGE \cite{wang2021bi} is composed of two such SAGEConv layers. GraphSAINT~\cite{graphsaintipdps19, graphsainticlr20} introduces a graph sampling-based inductive learning approach. The Bi-GraphSAINT~\cite{wang2021bi} model referred to in this paper comprises two GraphConv layers and a fully-connected layer. 

\subsection{Binary Neural Networks}
A bit-dot-product between two 0/1 bit-vectors can be computed as follows:
\[c = popc(a_{(b)} \ \& \ b_{(b)}),\]

\noindent where $\&$ denotes logic AND operation and $popc()$ is the population count function that counts the number of 1 bits. 

In binary neural networks (BNNs)~\cite{hubara2016binarized, courbariaux2016binarized, rastegari2016xnor, bulat2019xnor}, the binarization function is often defined as follows:
\[
    x_{(b)} = sign(x) 
\begin{cases}
    1,& \text{if } x\geq 0\\
    -1,& \text{otherwise}
\end{cases}
\]

The bit-dot-product operation of +1/-1 bit-vectors can be computed through:

\[c = n - 2 \times popc(a_{(b)} \oplus b_{(b)}) = 2 \times popc(a_{(b)} \odot b_{(b)}) - n\]
where $\oplus$ is exclusive-OR (XOR), $\odot$ is exclusive-NOR (XNOR), and $n$ is the \textit{bit-width} of $a_{(b)}$ and $b_{(b)}$.

\subsection{Bit-ops Intrinsics}\label{bitintrinsics}
Other than commonly-seen logical AND ($\&$), OR($|$), XOR ($\land$), negation ($\neg$), bit left-shifting ($\ll$) and right-shifting ($\gg$), GPUs are equipped with several integer intrinsics that can be used for efficient bit operations. \textsf{\_\_popc()} and \textsf{\_\_popcll()} allow fast bit-accumulation of a single 32-bit or 64-bit unsigned integer. \textsf{\_\_shfl\_sync()} is for register data exchange between threads in a warp \cite{li2018warp}. By indicating the thread lane to exchange, the intrinsic will receive register data from the designated thread lane. \textsf{\_\_ballot\_sync()} is a warp-voting intrinsic used for register data exchange between threads in a warp. Assuming all threads in a warp are active, it will return a bit-masked 32-bit unsigned integer showing the boolean evaluation result of the predicate-argument in each thread. It is the same as a clockwise transpose of a bit-column into a bit-row and can be useful for fast binarizing the full-precision values with a warp of threads. \textsf{\_\_brev()} and \textsf{\_\_brevll()} reverse the bit-order of a 32-bit and a 64-bit unsigned integer, respectively. 
When paired with \textsf{\_\_ballot\_sync()}, they can be helpful in rotating a bit-column 90$^{\circ}$ anti-clockwise into a bit-row \cite{li2019bstc}. Our implementation of efficient binary GNN is based on those intrinsics. As similar intrinsics can be found in other vendors' GPUs \cite{amd2022hip}, our BitGNN is portable to other GPU platforms.
\section{BitGNN}
Our design of BitGNN aims to achieve (i) efficiency: a GNN using BitGNN should enjoy a significantly higher speed; (ii) accuracy: a binary GNN that uses BitGNN should be able to get the same accuracy as the binary GNNs by prior approaches do; (iii) ease to use: with BitGNN, users should be able to convert a GNN into a binary GNN easily; (iv) flexible to tune: Different GNNs may require a different accuracy-speed objective; with BitGNN, users should be able to tune the tradeoff easily.

This section presents our design. The design of BitGNN consists of a set of novel abstractions, representations, algorithms, and optimizations. The final materialization is a library and utilities that users can use to build and tune their efficient binary GNNs easily. 

Two key components of a GNN are operations and tensors. Correspondingly, our creation of BitGNN centers around three key research questions: (R1) What should be the abstractions of common binary GNN operations? (R2) How should binary tensors be represented? (R3) How to assemble them into efficient ready-to-use libraries?  

\subsection{R1: What should be the abstractions of common binary GNN operations?}

GNNs have many kinds, and their binarized forms may consist of even more complexities and variations, depending on which part of the GNN is binarized for good accuracy-speed tradeoffs. For BitGNN to be easily applicable to various GNNs, it is hence important to abstract the common binary GNN operations into some building blocks. The result of such an abstraction must cover the most important operations of various binary GNNs, and at the same time, support their different needs in binarization.
To the best of our knowledge, no prior work has studied systematic abstractions of binary GNN operations.

\subsubsection{Complexities}
\label{sec:complexities}

\begin{figure*}
  \centering
  \includegraphics[trim={0cm 0cm 2cm 0cm},clip,width=.75\textwidth,height=.75\textheight,keepaspectratio]{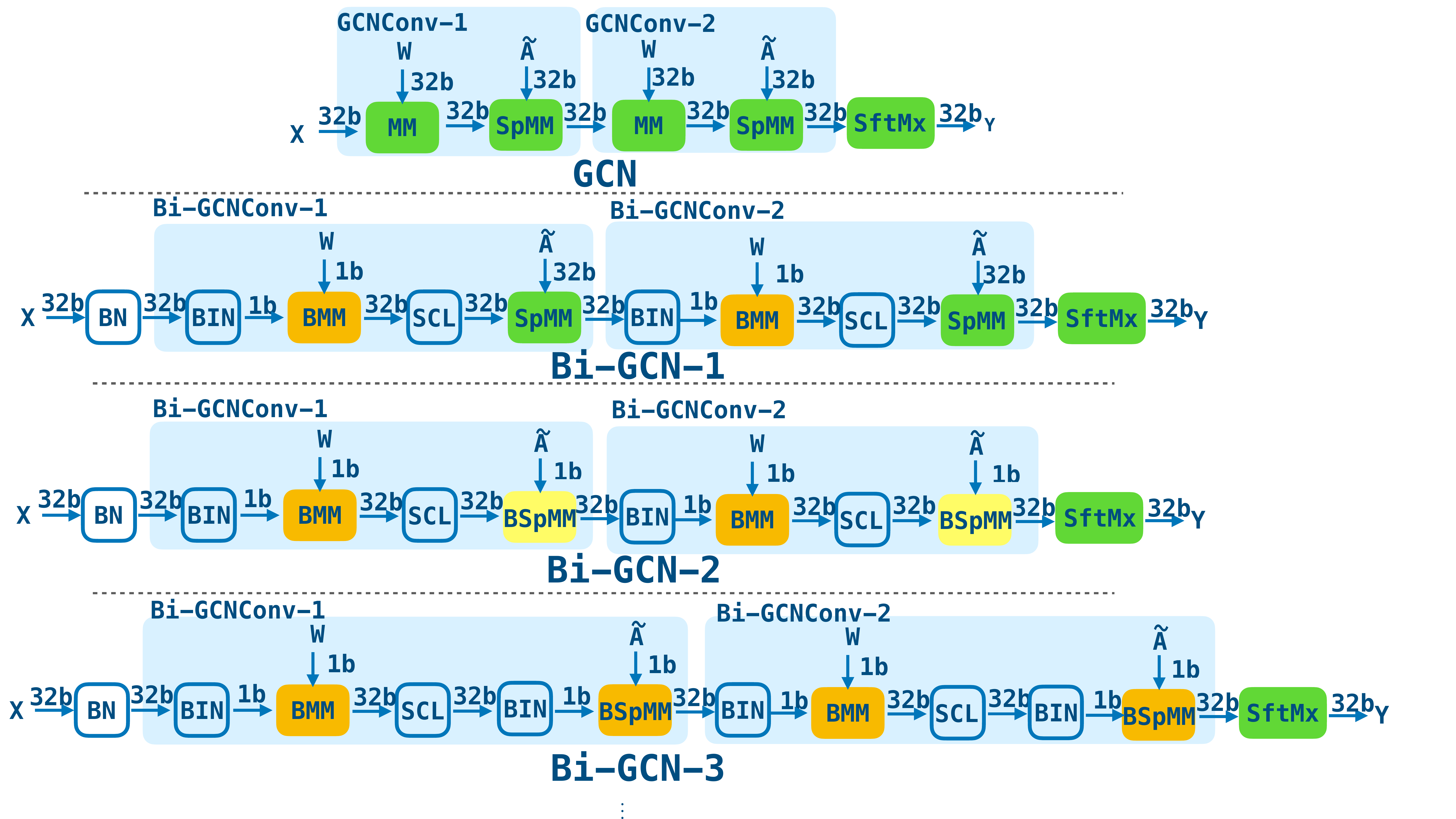} 
\caption{Illustration of the original full-precision GCN~\cite{kipf2017semi}, and several variants of binary GCN.}
  \label{fig:gcnExample}
\end{figure*}

To provide a deeper understanding of the complexities involved in designing binary Graph Neural Network (GNN) operations, this section takes a closer look at the binarization process using the Graph Convolutional Network (GCN) as an example.

The top graph in Figure~\ref{fig:gcnExample} illustrates the original GCN~\cite{kipf2017semi}. A forward pass of GCN involves two GCNConv layers followed by a softmax. Each GCNConv layer starts with a matrix-matrix multiplication (MM), where the input activation is multiplied by the weight matrix. The result at each graph node is then aggregated with its neighbor nodes' results, which is carried out by multiplication with the adjacency matrix of the graph. Because the adjacency matrix is typically sparse, the second matrix multiplication can use sparse matrix multiplication (SpMM).  

Below the original GCN in Figure~\ref{fig:gcnExample}, three variations of the binarized form of GCN are shown. The first one, Bi-GCN-1, is based on a previous work~\cite{wang2021bi}. It replaces the first MM with binary matrix multiplication (BMM) by binarizing the weight matrix offline. It also includes three extra operations - batch normalization (BN), a tensor binarization (BIN) before BMM, and a scaling operation (SCL) after the BMM to recover the scale of the results. The second MM is also replaced with BMM, and a BIN and an SCL operation are inserted before and after BMM, respectively.

The binarized form of Bi-GCN-1 focuses solely on the binarization of the two MM operations in the original GCN. However, other operations in the GCN could also be binarized, perhaps to varying degrees. For example, Bi-GCN-2 in Figure~\ref{fig:gcnExample} demonstrates a variant in which the two SpMM operations are "half" binarized - that is, the adjacency matrices are binarized, but the activation maps are not. In contrast, Bi-GCN-3 in Figure~\ref{fig:gcnExample} showcases a scenario where the two SpMM operations are fully binarized, with more auxiliary operations included. All three variations have their advantages and disadvantages: as the binarization increases, there is a potential for greater speedups, but also an increased risk of accuracy loss. It is evident that, besides these three variants, there can be numerous other variants, each corresponding to the combination of operations binarized to a certain degree.

Designing binary GNN operations becomes even more complex when the GNN contains additional operations, as illustrated by the original architectures of SAGE and SAINT depicted at the top of Figure~\ref{fig:abstraction}. Our analysis of numerous common GNNs and their binarized forms lead us to several observations:

(i) There can be multiple possible binarized forms for a given GNN, each involving numerous additional operations.

(ii) A simple one-on-one mapping is insufficient. After various binarizations, a single operation in the original GNN may be replaced with different sequences of operations. Therefore, creating only one abstraction for the binarized form of a specific original operation is inadequate.

(iii) There is a set of operations that constitute the core operations of binary GNNs. The variations depicted in Figure~\ref{fig:gcnExample}, for instance, are all composed of BN, BIN, BMM, SCL, SpMM, and softmax, despite their differences in specific connections and architectures. This pattern also emerges in other GNNs since most GNNs have MM and SpMM as their core operations.

\subsubsection{Two-level BitGNN Abstraction}~\label{sec:abstraction}

Based on our observations, we develop a two-level BitGNN abstraction and design them according to the following principles:

(1) Coverage: The abstraction must cover the most time-consuming core parts of common GNNs.

(2) Flexibility: The abstraction should support the need for different accuracy-speed tradeoffs.

(3) Efficiency: The abstraction should be mindful of the implications to computing efficiency.

(4) Ease of application: The abstraction should allow for easy adoption in GNN development so that a GNN can be easily revised into a binary GNN. Furthermore, it should support easy tuning of the binarization process to achieve different accuracy-speed objectives.

\begin{figure*}
  \centering
  \includegraphics[trim={0cm 0cm 0cm 0cm},clip,width=.75\textwidth,height=.75\textheight,keepaspectratio]{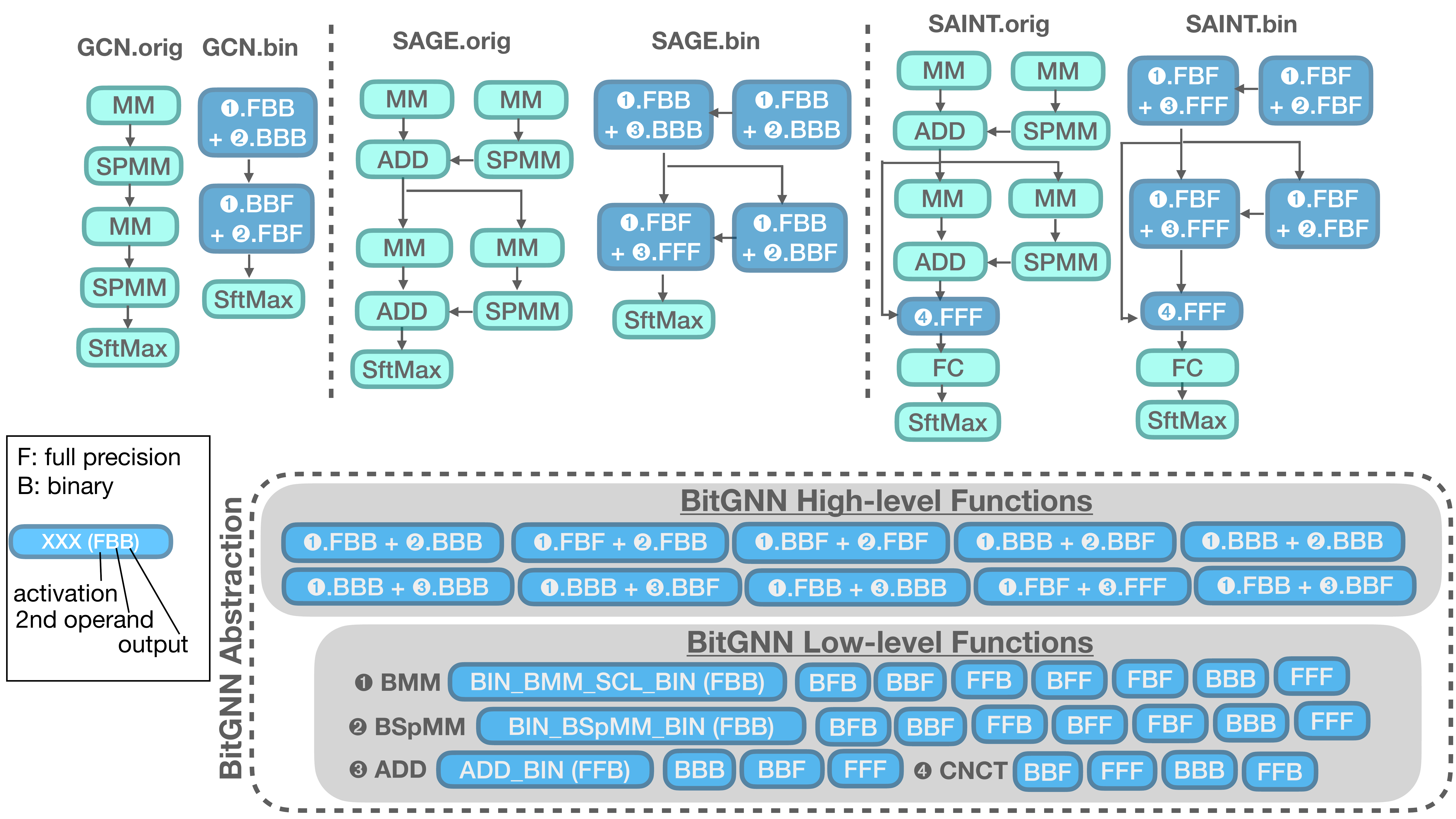}
  \caption{Core components of the two-level BitGNN abstraction (bottom) and its usage examples (top). At the low-level, BitGNN supports the core kernels in bit-ops, including \protect\circled{1} BMM (Bit-MatMul) \protect\circled{2} BSpMM (Bit-Sparse-Dense-MatMul)  \protect\circled{3}\protect\circled{4} Auxiliary Ops. At the high-level, BitGNN provides fused operators for binary or full-precision operand input/output. In the usage examples, "orig" and "bin" suffixes represent the original and binary forms of the GNNs, respectively.}
  \label{fig:abstraction}
\end{figure*}

The bottom of Figure~\ref{fig:abstraction} displays the core components of the abstraction. The low-level functions provide the primary building blocks, while the high-level functions offer options for drop-in replacement of the components in GNNs for binarization. The former offers flexibility, while the latter offers ease of use.

\paragraph{Low-level functions} The low-level functions are grouped into three categories. The first group focuses on the binarization of matrix multiplication (MM). In prior binary GNN materializations, one of the performance bottlenecks is the re-binarization of activation tensors in each layer. Our design of the functions takes this into consideration. We include seven BMM variants, each corresponding to a different combination of input activation, weight matrix, and output precisions, as shown at the bottom of Figure~\ref{fig:abstraction}. These precisions determine the auxiliary operands required (e.g., full-precision inputs require BIN before participation in the multiplication). In our design, these operands are included within the BMM functions to avoid invocation overhead and the need for complex kernel fusions.

One of our insights is that when BIN immediately follows SCL, the SCL becomes redundant and can be removed. The reason for this is that the scaling factor is always positive\footnote[2]{In Bi-GCN~\cite{wang2021bi}, for instance, the scaling factors are the row-wise and column-wise L1 normalization values.}, so the element-wise multiplication of the scaling factor will not affect the binarization result.

The second group of low-level functions is related to SpMM operations that are commonly used for multiplications involving adjacency matrices in GNNs~\cite{chen2022bitgraphblas}. Our design of binarized forms includes eight variants, as illustrated at the bottom of Figure~\ref{fig:abstraction}. The reasons for these variants include the accuracy-speed tradeoff, similar to the BMM case, and the special property of adjacency matrices. In GNN workloads, the connections between two nodes may or may not carry weights. In the former case, only 0/1 values are in the adjacency matrix, representing node connectivity. In the latter case, BSpMM can still be applied using a standard approach to binarize an adjacency matrix through factorization. Following the matrix multiplication, there will be a multiplication with a full-precision factorization vector.

The third group consists of auxiliary operations, such as the add operations for self-connectivity and concatenation operations. As binary operations may output results of different precisions, this group includes several variants. However, unlike other operations, mixed precisions of operands for these two operations are not meaningful in practice, and our design excludes those variants.

\noindent\fbox{\parbox{\columnwidth}{{\bf Notation:} for convenience, the following discussion uses a {\bf three-letter suffix} to distinguish the variants of an operator: For instance, BSpMM.FBB represents BSpMM that takes in a full-precision (F) matrix (1st operand) and a binary (B) matrix (2nd operand) as inputs and produces a binary (B) matrix.}}

\paragraph{High-level functions}
The low-level functions cover the most essential operations in binary GNNs and provide flexibility in meeting various needs. To facilitate adoption, we also design a set of high-level functions that enable simple drop-in replacement of key time-consuming operations in a GNN for binarization. Our research reveals that despite the wide range of GNNs available, their most time-consuming core operations are matrix multiplications (MM) and sparse matrix multiplications (SpMM). As a result, our high-level functions revolve around these operations and their combinations.

The high-level functions are organized into two groups, each combining multiple low-level functions to ease the adoption of binary GNNs. The first group is designed for a typical case where BSpMM immediately follows BMM, as seen in GCN illustrated in Figure~\ref{fig:abstraction}. The variants are determined by the activation tensor between the BMM and SpMM operations. It can be either in binarized form (B) or full-precision form (F). Consequently, we have BMM.FBB+BSpMM.BBB or BMM.FBF+BSpMM.FBB for the first GCNConv layer, and BMM.BBF+BSpMM.FBF or BMM.BBB+BSpMM.BBF for the second GCNConv layer. The second group is designed for another common case where some auxiliary operations follow MM, as seen in the architecture of SAGE in Figure~\ref{fig:abstraction}. The auxiliary operator comes from self-connectivity, for which we can use either ADD.BBF or ADD.FFF to merge the activation output at the end of the first SAGEConv layer.

The top of Figure~\ref{fig:abstraction} demonstrates how the three GNNs can be easily converted into binary GNNs by replacing the common operation sequences with the corresponding high-level functions. Low-level functions with small latency, such as ADD, are fused with BMMs or BSpMMs, while high-level functions are optimized through inter-layer fusions between BMMs and BSpMMs using cooperative kernel launch to improve thread block synchronization. Furthermore, the high-level functions enable easy tuning of binary GNNs for accuracy-speed tradeoffs. For instance, the user can replace the two MM-SpMMs in GCN.bin shown in Figure~\ref{fig:abstraction} with other options in the high-level function set. As long as the output precision of a predecessor block matches the input precision of its successor, the correctness of types is guaranteed, while the accuracy and speed may vary.

To summarize, BitGNN's design of high-level and low-level abstractions provides comprehensive coverage of the decision space for binary GNN architecture exploitation. At the low-level, the design addresses the various precision requirements of the core operations while minimizing invocation overhead. At the high-level, the design avoids unnecessary rebinarization, offers flexibility for tuning accuracy-speed tradeoffs, and enables straightforward drop-in replacement for converting GNNs to binary GNNs.

\subsection{R2: How to represent binarized tensors?}\label{solutions}

The representation of binarized tensors plays a crucial role in determining the efficiency of memory usage and data access speed of binarized GNNs. However, existing binary GNN implementations such as those in~\cite{hubara2016binarized, courbariaux2016binarized, rastegari2016xnor, bulat2019xnor} do not truly store activation or weight tensors in bits, nor do they consider storing the graph in bits.

In conventional DNNs, some work has explored the representation of binary tensors and proposed various representations, as seen in~\cite{li2019bstc, li2020accelerating}. Nevertheless, two special complexities in GNNs call for innovative solutions. In the following, we will focus on discussing each of these complexities and the solutions we propose.

\subsubsection{Granularity Dilemma}

The first complexity is about the granularity of representations. 

\paragraph{Dilemma}

There are many representations proposed for sparse matrices, and one that shows a particularly good fit for adjacency matrices is block-based representation~\cite{chen2022bitgraphblas}. In this representation, the matrix is viewed as a composition of many k-by-k blocks, with each block's content stored in a dense format. The block-level representation uses a sparse format (e.g., CSR) with all-zero blocks being ignored.

This representation, however, faces a dilemma with block size. The smaller the block size, the fewer zeros will be included in the representation, but it also leads to more reduction operations among blocks. A smaller block is also less friendly to the massive parallelism of the GPU.

\paragraph{Solution}

To address the dilemma of block size, we propose {\em fine-representing dynamic-coarsening (FRDC)}. This approach utilizes a fine-grained representation to achieve high space efficiency and on-chip coarsening to attain high time efficiency. Specifically, FRDC uses a 4\texttimes{}4 block size for representation. The choice of 4\texttimes{}4 is motivated by the need to use a fine-grained representation to store matrices while using online stitched coarse-grained representation for computations. This approach saves storage and memory space while better exploiting bit-level parallelism.

Using larger block sizes such as 4\texttimes{}8 or 8\texttimes{}4 doubles memory usage, while smaller tiles lead to increased dynamic stitching overhead. Our empirical results demonstrate that 4\texttimes{}4 is a suitable block size. During computations, the 4\texttimes{}4 bit-blocks containing 16 bits are transferred from GPU DRAM to shared memory/registers. The BitGNN kernel then assembles the 4\texttimes{}4 bit-blocks into 32-bit aligned words on shared memory and efficiently processes them using word-level bit manipulation intrinsics (see Section~\ref{bitintrinsics}). By using this approach, we can maintain the space benefits of the fine-grained representation while achieving maximal parallelism.

\subsubsection{Inconsistent Value Ranges}\label{sec:trinary}

The second special complexity arises from the different value ranges of tensors after binarization. 

One common situation is the inconsistency between adjacency matrices and activation maps. While adjacency matrices use 1/0 to denote edge connectivity, the basic operation is a 0/1 dot-product. On the other hand, binary neural network activation uses the +1/-1 dot product. This inconsistency presents a challenge when conducting multiplications between them efficiently. However, previous studies have not addressed this issue.

In this work, we propose three methods to reconcile the incompatibility. Let $a$ represent a 0/1-based bit-vector of the adjacency matrix, and $b_{org}$ be a +1/-1-based bit-vector from the activation matrix. Both $a$ and $b_{org}$ are represented in 0/1 bits in memory, but for the sake of explanation, we introduce $b$ to represent the in-memory 0/1 representation of $b_{org}$, where 1 corresponds to +1 and 0 corresponds to -1. We now describe the three methods to compute the dot product between $a$ and $b_{org}$.

(1) If-else evaluation on $a$'s nonzeros: This method examines each bit of $a$. If the value is 1, we add the corresponding bit in $b$ (1 $\rightarrow$ 1, 0 $\rightarrow$ -1) to the temporary sum; otherwise, we skip the bit value in $b$.

(2) $popc(a\&b)-popc(a\&\neg b)$: $popc(a\&b)$ accumulates the total number of 1s of $a \times b$. $popc(a\&\neg b)$ ($\neg b$ represents the complement of $b$) gives the total number of -1s in $b_{org}$. Subtracting the two results provides the dot product of $a$ and $b_{org}$.

(3) $2 \times popc(a\&b)-popc(a)$: The transformation from +1/-1 to 1/0 involves adding 1 and then dividing by 2 (i.e., right shift). Thus, $b=(b_{org}+1)/2$. Consequently, $b_{org}=2 \times b-1$. Since $popc(a\&b)$ is the bit-dot-product of $a$ and $b$, we can compute the dot-product of $a$ and $b_{org}$ as $2 \times popc(a\&b)-popc(a)$.

If the activation values are in bits, Solution-2 and Solution-3 are more efficient than Solution-1. However, if the activation values are in other forms, Solution-1 may provide better performance. Section~\ref{sec:tune} provides additional discussions on selecting the appropriate method.

\subsection{R3: How to realize the potential in coding?}

The third research question pertains to the effective materialization of BitGNN abstractions to fully leverage the efficiency benefits of binary representations of tensors. 

Previous works on binary graph neural networks, such as \cite{wang2021bi, bahri2021binary, wang2021binarized}, have only simulated binarization using the \textsf{sign()} function and FP32 multiplication, without treating values as bits in the implemented operations.

To address this issue, our proposed solution meticulously explores the bit manipulation intrinsics offered by modern GPUs to fully unlock the performance potential of BitGNN. In BitGNN, the core operations are binary matrix multiplication (BMM) and binary sparse matrix multiplication (BSpMM), as illustrated in Figure~\ref{fig:abstraction}. Both operations have several variants. While previous studies have examined BMM~\cite{li2019bstc,li2020accelerating}, the ones in BitGNN build upon prior work while taking into account various precisions and the fusion of auxiliary operations. However, BSpMM has not been previously studied. Thus, the remainder of this discussion focuses on BSpMM.

\subsubsection{Design Principles} 

The BSpMM kernels in BitGNN that use FRDC-based tensors are designed based on the following principles:

(1) \textit{Warp-based workload partition}: We partition each workload unit of a node into one or more warps and utilize fast intra-warp communication~\cite{ben2016fast} on GPUs to its fullest potential.

(2) \textit{Maximizing bit-level parallelism}: This addresses the granularity dilemma mentioned earlier and maximizes the edge traversal throughput (measured in the number of traversed edges per second (TEPS)). We ensure that the small bit-blocks in sparse graphs are ultimately manipulated in a word-aligned fashion.

(3) \textit{Maximizing bit-tensor load \& store efficiency}: SpMM is often memory-bound~\cite{huang2020gespmm, rahman2020fusedmm}, so we must carefully use the memory hierarchy for bit-tensor manipulations to ensure maximum efficiency.
\subsubsection{Implementations}
\begin{figure*}
    \centering
    \includegraphics[trim={0cm 0.5cm 0cm 0.5cm},clip,width=.8\textwidth,height=.8\textheight,keepaspectratio]{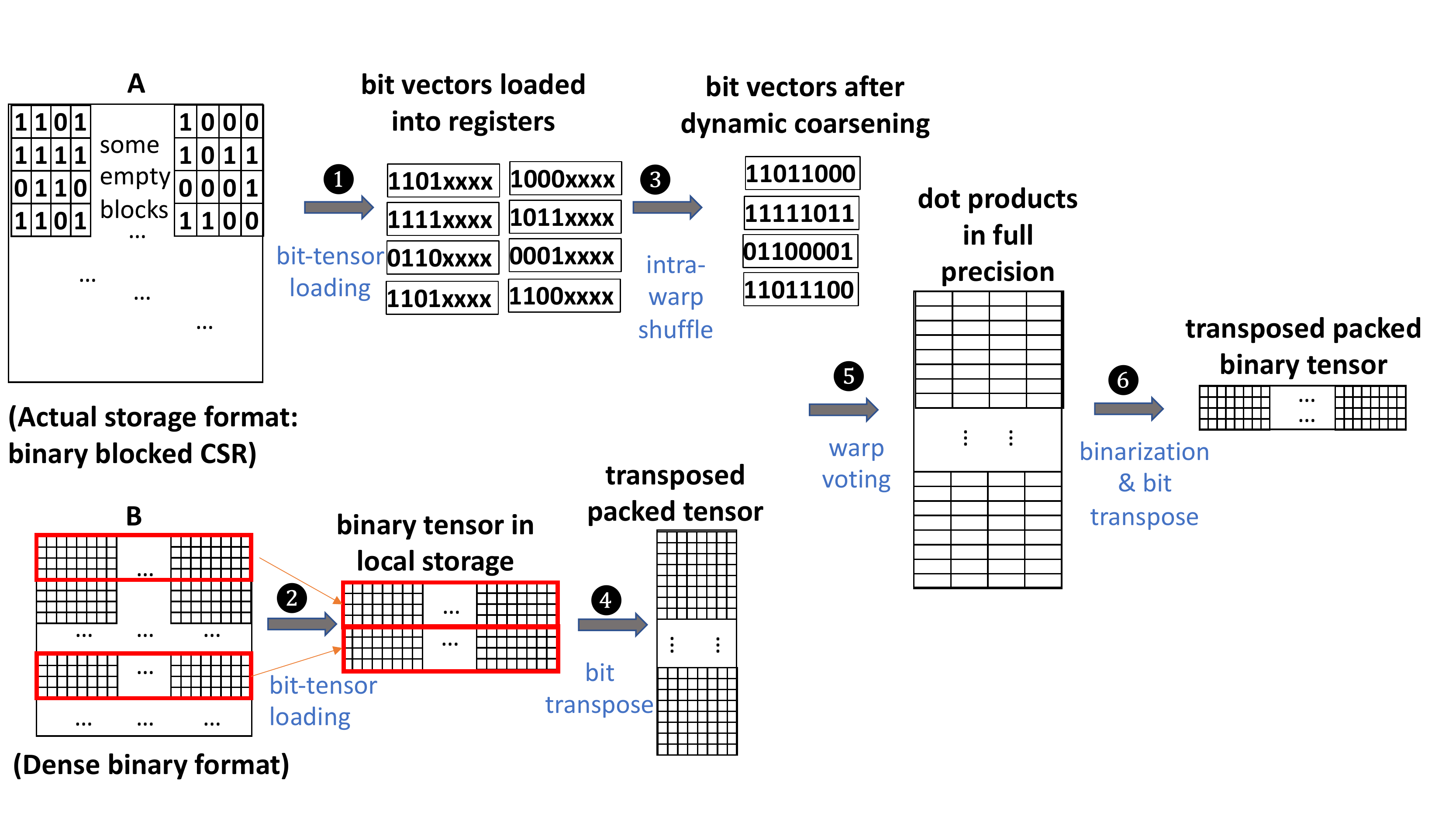}
    \caption{Illustration of the procedure of BSpMM.BBB using FDRC (Fine-Representing Dynamic-Coarsening).}
    \label{fig:bspmm}
\end{figure*}

Following the three principles mentioned earlier, we implement the BitGNN abstractions using the FRDC scheme and bit-manipulation intrinsics. We will explain the implementation of BSpMM.BBB in detail as an example.

\begin{algorithm}[!ht]
\DontPrintSemicolon
\KwInput{AdjMat: RowPtr, ColInd, BitTiles, Tile-row ID $r$; InActMat$_{(b)}$;} 
\KwOutput{OutActMat$_{(b)}$}
$NT \gets $ RowPtr[$r$+1]-RowPtr[$r$]; $TS \gets \frac{SIMD\_unit}{tile\_dim}$

\For{each tile set $t_{s}$ where $s$ = 1 to $\lceil\frac{NT}{TS}\rceil$}
{
    \textbf{register} A $\gets$ BitTiles[$s$] for $s$ = 0 to $TS$-1
    
    \textbf{register} B $\gets$ InActMat$_{(b)}$[$k$] for $k$ = ColInd[$s$] $\times$ $tileDim$ + \{0,1,2,...,$tileDim$-1\} and $s$ = 0 to $TS$-1
    
    \textbf{register} a $\gets$ a $|$ \codefont{shfl(}A, $s$$\times$$tile\_dim$+$laneid$\codefont{)} $\ll$ (4$\times$($TS$-1-$s$)) for $s$ = 0 to $TS$-1 \tcp{do bit-concatenate}
    
    \textbf{shared memory} b[$warpid$$\times$$\lceil\frac{32}{W}\rceil$+$k$] $\gets$ \codefont{brev(ballot(}(B$\gg$(31-$k$))\&0x1\codefont{))} for $k$ = 0 to $SIMD\_unit$-1 \tcp{do bit-transpose}
    
    \textbf{register} c[$n$] $\gets$ \codefont{popc(shfl(}a,$n$\codefont{)}\&b[$warpid$$\times$$\lceil\frac{32}{W}\rceil$+$laneid$]\codefont{)} - \codefont{popc(shfl(}a,$n$\codefont{)}\&$\neg$b[$warpid$$\times$$\lceil\frac{32}{W}\rceil$+$laneid$]\codefont{)} for $n$ = 0 to $tileDim$-1 \tcp{compute bit-dot-product}
}
\textbf{register} rs[$n$] $\gets$ \codefont{brev(ballot(}c[$n$]$\geq$0\codefont{))} for $n$ = 0 to $tileDim$-1

OutActMat$_{(b)}$[$r$$\times$$tileDim$+$n$] $\gets$ rs[$n$] for $n$ = 0 to $tileDim$-1

\caption{BSpMM.BBB}\label{alg-bbspmm}
\end{algorithm}

\noindent \textbf{BSpMM.BBB} Algorithm: In Algorithm~\ref{alg-bbspmm}, we present the implementation of the BSpMM.BBB kernel using the FRDC technique. The binary adjacency matrix is represented using \textsf{RowPtr}, \textsf{ColInd}, and \textsf{BitTiles (BitBlock)}\footnote[3]{We use the terms "tile" and "block" interchangeably in the paragraphs.}. The algorithm is parallelized through the tile (block) row dimension, and we present tile (block) row \textsf{r}'s workload.

In Line 1, each tile-row's total number of tiles ($NT$) is divided into several tilesets. The size of a tileset ($TS$) is defined as the number of bit-row required to concatenate into a \textit{SIMD\_unit}, which is equal to $\frac{SIMD\_unit}{tile\_dimension}$. For example, when the adjacency matrix is stored in the block-$4 \times 4$ format with $4 \times 4$ non-empty tiles, a non-empty tile contains four nibbles (4-bit). When we use 32 as a \textit{SIMD\_unit}, the tileset size ($TS$) equals 8. Consequently, in Line 2, the kernel will process 8 bit-tiles in each iteration. A tileset ($TS$) is the processing unit of an iteration, and the workload of each tile-row requires $\lceil\frac{NT}{TS}\rceil$ iterations to accomplish.

In this scenario, both the adjacency matrix ($A$ in Figure~\ref{fig:bspmm}) and input activation ($B$ in Figure~\ref{fig:bspmm}) are binary, and the resulting output should also be binary. As shown in Figure~\ref{fig:bspmm}, the algorithm includes the following steps:

Step \circled{1} \& \circled{2}: \underline{Warp-based workload partitioning and loading}. To efficiently utilize the GPU architecture, each warp of 32 threads is mapped to the computation of a row of 4\texttimes{}4 bit-tiles (i.e., 4 nodes when representing the graph in bits). Within each warp, the row of bit-tiles is further partitioned into multiple tile groups, where each tile group contains 8 bit-blocks. For example, if there are ten bit-tiles, there will be two tile groups. The first tile group holds the 0th-7th tiles of that 4\texttimes{}4 tile row, and the second tile group holds the 8th and 9th tiles, with the remaining 8 slots padded with zeros.

Next, each iteration processes a tile group as follows: the 32 threads in the warp cooperatively load the 32 uchars\footnote[4]{We store 4-bit unit as an uchar (1 byte) in our original implementation. Storing them in a smaller representation (i.e., int4) is also possible in newer GPU architectures.} from 8 bit-tiles of the sparse bit matrix $A$ from global memory into registers. Then, the 32 threads load the 8 segments of the input activation $B$ corresponding to the loaded tiles of $A$ from global memory into shared memory, forming a consecutive layout, with each word used as a 32-bit vector.

Step \circled{3} \underline{On-the-fly assembly via bit-concatenation}. In this step, dynamic coarsening is performed to ensure that the bit-tiles of $A$ are 32-bit aligned. Each of the 0th-3rd threads uses the intra-warp communication intrinsic (\textsf{shfl\_sync()}) and bit shifting to quickly concatenate the neighbors of each node (in the form of bit-vectors) into a 32-bit unsigned bit-vector (only 8 bits are shown in our illustration in Figure~\ref{fig:bspmm}).

Step \circled{4} \underline{Bit-transpose for column-wise coalesced access}. This step involves transposing the column-major packing of $B$ into row-major. This way, when computing the bit-dot-product with $A$, the bits can be fetched as bit-columns, and the bit-level parallelism can be fully employed. The transpose is accomplished through bit shifting and masking on the bit-vector of $B$. The resulting values are then evaluated and gathered using the intra-warp voting intrinsic, \textsf{ballot\_sync()}. Finally, the transposition is done using the \textsf{brev()} intrinsic.

Step \circled{5} \underline{Trinary-valued bit-dot-product}. In this step, the algorithm computes the bit-dot-product between the 0/1 bit-vectors of A and the +1/-1 bit-vectors of B. (``Trinary'' for 0/1/-1.) The loop iterates 4 times (once for each node), working on a 32-bit bit-vector from $A$. Each of the 32 threads is responsible for computing 1/32 of the bit-dot-product. The products are stored in registers as full-precision values.

Step \circled{6} \underline{Bit-tensor store}. In this step, the temporary full-precision products are packed into the bit format by element-wise evaluation of whether the product is greater than or equal to 0, which is the binarization process of binary activation tensor. The 32 threads in a warp cooperatively evaluate the full-precision values using the intra-warp voting intrinsic (i.e., \textsf{ballot\_sync()}). The result is then anti-clockwise transposed by the \textsf{brev()} intrinsic to complete the bit-vector packing. The output produces 32-bit unsigned bit-vectors of the activation map.

\noindent \textbf{Other variants} The implementation of other variants of BSpMM follows a similar approach, using bit-level intrinsics for most computations with some differences in handling different precisions. For example, BSpMM.FBB loads the activation tensor differently. In Step \circled{2} (bit-tensor load), it requires 4 warps to cooperatively load all the full-precision values. Each thread in a warp is responsible for loading one of the dimensions of the feature vectors of 32 neighbor nodes. Therefore, Step \circled{4} (Bit-transpose for column-wise coalesce access) is not required. More details can be found in the supplementary material~\cite{bitgnnapendix}.

\subsection{Tuning Utilities and Other Implementation Details}\label{sec:tune}

To launch our end-to-end model, which contains multiple kernels, we use \textsf{cudaLaunchCooperativeKernel()} to enable global synchronization between a specific set of thread blocks~\cite{cudaprogrammingguide2022}. We first calculate the maximum number of blocks that fit each streaming multiprocessor (SM) using \textsf{cudaOccupancyMaxActiveBlocksPerMultiprocessor()} and then use 1024 threads per thread block to allow the maximum number of warps to execute on each SM. This ensures that we fully utilize the parallelism available on the GPU and achieve optimal performance.

BitGNN's tuning utilities provide a convenient way to optimize binary GNNs. As described in the previous sections, BitGNN offers a range of variants to accommodate the different precision requirements of GNN inputs, weights, and outputs. The tuning utilities allow for an auto-tuning run to replace BitGNN function calls with other variants and measure their performance. The type correctness is guaranteed by ensuring that the precision of the output of a predecessor matches that of the input of its successor operation. Furthermore, the tuning utilities enable easy selection of the best solution for reconciling the inconsistency of binary value ranges discussed in Section~\ref{sec:trinary}. Predictors can be developed to anticipate the optimal variant for a particular GNN and graph. Similar predictors have been extensively studied in other contexts, such as sparse matrix storage formats~\cite{zhao2018}.
\section{Evaluation} \label{sec-evaluation}
This section reports the performance of BitGNN. We focus on the following questions: (i) How much speedup can it bring to GNN inferences? (ii) How much memory space can it same? (iii) What are the accuracy-speed tradeoffs when using BitGNN?

\subsection{Methodology}

\begin{table*}
\caption{GPUs used for evaluation. Arch refers to GPU architecture generation. CC refers to compute capability. SMs refer to the number of streaming multiprocessors in the GPU. Thrds refer to threads. Shared refers to share memory size. Reg refers to number of registers. Note, "/Block" and "/Thrd" imply the maximum resources per thread block and per thread, respectively.}
\label{tab-gpu-memory}
\resizebox{2\columnwidth}{!}{
\begin{tabular}{|c|c|c|c|c|c|c|c|c|c|c|c|c|c|c|c|}
\hline
\textbf{GPU} & \textbf{Arch} & \textbf{CC} & \textbf{SMs} & \textbf{DRAM} & \textbf{\begin{tabular}[c]{@{}c@{}}Memory \\ Bandwidth\end{tabular}} & \textbf{\begin{tabular}[c]{@{}c@{}}L1 Cache \\ Size/SM\end{tabular}} & \textbf{\begin{tabular}[c]{@{}c@{}}L2 Cache \\ Size\end{tabular}} & \textbf{\begin{tabular}[c]{@{}c@{}}Warps\\ /SM\end{tabular}} & \textbf{\begin{tabular}[c]{@{}c@{}}Blocks\\ /SM\end{tabular}} & \textbf{\begin{tabular}[c]{@{}c@{}}Thrds\\ /SM\end{tabular}} & \textbf{\begin{tabular}[c]{@{}c@{}}Shared\\ /SM\end{tabular}} & \textbf{\begin{tabular}[c]{@{}c@{}}Shared\\ /Block\end{tabular}} & \textbf{\begin{tabular}[c]{@{}c@{}}Reg\\ /SM\end{tabular}} & \textbf{\begin{tabular}[c]{@{}c@{}}Reg\\ /Block\end{tabular}} & \textbf{\begin{tabular}[c]{@{}c@{}}Reg\\ /Thrd\end{tabular}} \\ \hline
GTX 1080     & Pascal        & 6.0          & 20           & 8GB           & 320GB/s                                                               & 48KB                                                                 & 4096KB                                                            & 64                                                           & 32                                                            & 2048                                                         & 64KB                                                          & 48KB                                                             & 64K                                                        & 64K                                                           & 255                                                          \\ \hline
TITAN V      & Volta         & 7.0          & 80           & 12GB          & 653GB/s                                                               & 96KB                                                                 & 4608KB                                                            & 64                                                           & 32                                                            & 2048                                                         & 96KB                                                          & 96KB                                                             & 64K                                                        & 64K                                                           & 255                                                          \\ \hline
RTX 3060 Ti  & Ampere        & 8.0          & 38           & 8GB           & 448GB/s                                                               & 128KB                                                                & 4096KB                                                               & 64                                                           & 32                                                            & 2048                                                         & 164KB                                                         & 164KB                                                            & 64K                                                        & 64K                                                           & 255                                                          \\ \hline
\end{tabular}
}
\end{table*}
\begin{table}
\caption{Graph datasets used for evaluation.}
\label{tab-datasets}
\begin{tabular}{ccccc}
\hline
\textbf{Dataset} & \textbf{\#Nodes} & \textbf{\#Edges} & \textbf{\#Features} & \textbf{\#Classes} \\ \hline
\textbf{Cora }            & 2,708            & 13,264           & 1,433               & 7                  \\
\textbf{Pubmed}           & 19,717            & 108,356           & 500                 & 3                  \\
\textbf{Citeseer}         & 3,327            & 12,431            & 3,703               & 6                  \\
\textbf{Flickr}           & 89,250           & 899,756          & 500                 & 7                  \\
\textbf{Reddit}          & 232,965          & 114,615,892      & 602                 & 41                 \\
\hline
\end{tabular}
\end{table}

\noindent\underline{\textbf{GPU Environment:}} We evaluate the performance of BitGNN on three NVIDIA GPUs of different architecture generations: GTX 1080 (Pascal), Titan V (Volta), and RTX 3060 Ti (Ampere). Their features are summarized in Table~\ref{tab-gpu-memory}. The CUDA version is 11.0. For all experiments, we use the average values of 10 repeated executions.

\noindent\underline{\textbf{Datasets:}} Table~\ref{tab-datasets} shows the graphs used for the evaluation. These graphs are commonly used for existing GNN research~\cite{hamilton2017inductive, yang2016revisiting, graphsaintipdps19, graphsainticlr20, hu2020open, shchur2018pitfalls, zitnik2017predicting} and are also the graphs used in prior binary GNN work Bi-GCN~\cite{wang2021bi}, making direct comparisons possible.

\noindent\underline{\textbf{Versions to compare:}} We compare our BitGNN versions (of different previsions) with Bi-GCN~\cite{wang2021bi}. As the representative of the state-of-the-art binary GNNs---which all focus on binarization algorithms rather than optimized executions, Bi-GCN uses the full-precision representation of values even though it binarizes the operations in GNNs. Meanwhile, to provide a reference point, we report the performance and accuracy of the original GNNs without binarization. Both the original and Bi-GCN versions are in PyG~\cite{fey2019fast}.

\noindent\underline{\textbf{GNNs:}} We use four GNNs in our experiments: transductive-GCN~\cite{kipf2017semi}, inductive-GCN~\cite{hamilton2017inductive}, GraphSAGE~\cite{hamilton2017inductive}, and GraphSAINT~\cite{graphsaintipdps19, graphsainticlr20}. We choose these GNNs because they are the most popular GNNs, and are the GNNs used in the prior binary GNN study~\cite{wang2021bi} that we compare with.

\subsection{Results and Analysis}
\begin{table}
\caption{Evaluation results of transductive GCNs. FP32 is the vanilla GCN~\cite{kipf2017semi}. "FP32 (S)" refers to the PyG scatter-gather abstraction with maximal batch size. "FP32 (T)" refers to the PyG SpMM tensor abstraction. Bi-GCN~\cite{wang2021bi} is the state-of-the-art binary GCN implementation. We compare the version that binarizes both activations and weights in the MM operation. "Ours (full)" refers to the {\em full-precision aggregation}. "Ours (bin)" refers to the {\em binary aggregation} version that uses MM.FBB + BSpMM.BBB for the first layer and MM.BBF + BSpMM.FBF for
the second layer. "Peak Mem" refers to peak memory usage during inference. "Acc" refers to accuracy in percentage (\%).}
\label{tab-transductive}
\centering
\resizebox{\columnwidth}{!}{
\begin{tabular}{ccccccc}
\hline
\multirow{2}{*}{\textbf{Dataset}}  & \multirow{2}{*}{\textbf{Model}} & \multirow{2}{*}{\textbf{\begin{tabular}[c]{@{}c@{}}Peak\\ Mem (B)\end{tabular}}} & \multirow{2}{*}{\textbf{\begin{tabular}[c]{@{}c@{}}Acc\\ (\%)\end{tabular}}} & \multicolumn{3}{c}{\textbf{End-to-end Time (ms)}}       \\ \cline{5-7} 
                                   &                                 &                                                                                  &                                                                              & \textbf{GTX1080} & \textbf{TitanV} & \textbf{RTX3060Ti} \\ \hline
\multirow{5}{*}{\textbf{Cora}}     & FP32 (S)~\cite{kipf2017semi}    & 16.73M                                                                           & 81.4$\pm$0.4                                                                   & 1.25$\pm$0.01      & 0.91$\pm$0.03     & 1.00$\pm$0.01        \\
                                   & FP32 (T)~\cite{kipf2017semi}    & 16.73M                                                                           & 81.4$\pm$0.4                                                                   & 1.34$\pm$0.02      & 1.15$\pm$0.05     & 1.16$\pm$0.03        \\
                                   & Bi-GCN~\cite{wang2021bi}        & 16.73M                                                                           & 81.2$\pm$0.8                                                                   & 1.84$\pm$0.04      & 1.91$\pm$0.02     & 1.57$\pm$0.02        \\
                                   & \textbf{Ours (full)}            & 1.37M                                                                            & 81.2$\pm$0.8                                                                   & 0.51$\pm$0.00      & 0.37$\pm$0.07     & 0.38$\pm$0.02        \\
                                   & \textbf{Ours (bin)}             & 0.73M                                                                            & 81.2$\pm$1.0                                                                   & 0.32$\pm$0.04      & 0.28$\pm$0.08     & 0.24$\pm$0.04        \\ \hline
\multirow{5}{*}{\textbf{PubMed}}   & FP32 (S)                        & 48.71M                                                                           & 79.0$\pm$0.3                                                                   & 2.86$\pm$0.04      & 1.51$\pm$0.02     & 1.90$\pm$0.02        \\
                                   & FP32 (T)                        & 48.71M                                                                           & 79.0$\pm$0.3                                                                   & 2.26$\pm$0.01      & 1.49$\pm$0.02     & 1.77$\pm$0.01        \\
                                   & Bi-GCN                          & 48.71M                                                                           & 78.2$\pm$1.0                                                                   & 2.81$\pm$0.07      & 2.23$\pm$0.09     & 1.52$\pm$0.01        \\
                                   & \textbf{Ours (full)}            & 7.31M                                                                            & 78.2$\pm$1.0                                                                   & 2.29$\pm$0.05      & 0.84$\pm$0.05     & 1.14$\pm$0.00        \\
                                   & \textbf{Ours (bin)}             & 2.65M                                                                            & 78.1$\pm$1.1                                                                   & 0.74$\pm$0.04      & 0.33$\pm$0.05     & 0.44$\pm$0.01        \\ \hline
\multirow{5}{*}{\textbf{CiteSeer}} & FP32 (S)                        & 49.78M                                                                           & 70.9$\pm$0.5                                                                   & 2.60$\pm$0.03      & 1.43$\pm$0.04     & 1.74$\pm$0.01        \\
                                   & FP32 (T)                        & 49.78M                                                                           & 70.9$\pm$0.5                                                                   & 2.70$\pm$0.01      & 1.56$\pm$0.01     & 1.88$\pm$0.05        \\
                                   & Bi-GCN                          & 49.78M                                                                           & 68.8$\pm$0.9                                                                   & 2.44$\pm$0.00      & 2.22$\pm$0.04     & 1.33$\pm$0.09        \\
                                   & \textbf{Ours (full)}            & 2.56M                                                                            & 68.8$\pm$0.9                                                                   & 0.80$\pm$0.04      & 0.58$\pm$0.02     & 0.50$\pm$0.03        \\
                                   & \textbf{Ours (bin)}             & 1.77M                                                                            & 68.7$\pm$0.4                                                                   & 0.63$\pm$0.03      & 0.46$\pm$0.01     & 0.42$\pm$0.00        \\ \hline
\end{tabular}
}
\end{table}

Tables~\ref{tab-transductive}, \ref{tab-inductive}, and~\ref{tab-saint} report the speed (at inference time), accuracy, and space usage of the baselines and the versions of our BitGNN. 

Before diving into the results, it is worth mentioning that GNN learning is of two kinds: transductive and inductive learning. In transductive learning, the entire data graph is observed in the learning process, with some nodes labeled and others not. The GNN algorithm tries to iteratively refine the labels of those unlabeled nodes by learning from the entire graph structure and those labeled nodes. In inductive learning, the GNN algorithm learns from some sampled graphs and then tries to predict the labels of the nodes on other sampled graphs. 

Among the four GNNs used in prior research~\cite{kipf2017semi, hamilton2017inductive, graphsaintipdps19, graphsainticlr20}, {\em transductive-GCN} is for transductive learning, and the other three GNNs are for inductive learning. The prior work applies {\em transductive-GCN} to only small datasets (\texttt{Cora, PubMed, CiteSeer}) because large graphs are difficult to be loaded and processed in memory as an entirety as required by transductive GNNs. For the large datasets (\texttt{Flickr} and \texttt{Reddit}), the prior work applies the three inductive GNNs. To allow head-to-head comparisons, our experiments follow the same practice. 

The proposed flexible programming abstraction (Section~\ref{sec:abstraction}) can easily facilitate the user's binary network development. With it, we tune the various binarization precisions and identify some model variants that perform relatively well: \textit{full-precision aggregation} (the same as in Bi-GCN~\cite{wang2021bi}) and \textit{binary aggregation} (Use MM.FBB + BSpMM.BBB for the first layer and MM.BBF + BSpMM.FBF for the second layer). They are represented as {\em Ours (full)} and {\em Ours (bin)} in the result tables. 

\noindent\underline{\textbf{Transductive case:}} From Table~\ref{tab-transductive}, we have the following observations. (i) Both versions of BitGNN can keep most of the accuracy. The largest loss of the average accuracy is 0.2\%, 0.9\%, and 2.2\% on the three datasets, respectively. The largest accuracy loss (2.2\%) happens on {\em CiteSeer}, where the normalization of edge weights has a more significant impact.  (ii) Both versions can save memory usage significantly, 7--19X by {\em Ours (full)} and 18-28X by {\em Ours (bin)}. (iii) Both versions consistently show significant speedups across GPU models: 1.2-5X by {\em Ours (full)} and 4-7X by {\em Ours (bin)}. (iv) Because the previous binarized GNN (Bi-GCN) does not use a single bit but a word for a binarized value, it does not save space or time compared to the original full-precision GNNs. (v) Our more complete binarized version ({\em Ours (bin)}) achieves 2-3X extra space savings and 1.5-3X extra speedups than our version ({\em Ours (full)}) that uses full-precision aggregation, while the extra accuracy loss is only 0.1\%. (vi) All those benefits are largely consistent across the datasets. One exception is the performance of {\em Ours (full)} on \texttt{PubMed}, where, {\em Ours (full)} shows modest speedups on GTX1080 and RTX3060Ti over the original version while {\em Ours (bin)} still shows substantial speedups. We attribute this to the relatively larger \textsf{num\_of\_nodes} that introduces large performance gaps between BSpMM.BBB and BSpMM.FBB for the first layer. 

\noindent\underline{\textbf{Inductive case:}} Tables~\ref{tab-inductive} and~\ref{tab-saint} report the results of the three inductive GNNs on the two large datasets. The benefits are also significant in space and time, but there are some differences from those in the transductive case. We note the following points. (i) For inductive learning, the previous work~\cite{hamilton2017inductive} uses the F1-micro score rather than accuracy to measure the quality of the inferences. The F1-micro score is a combined metric of precision and recall~\cite{hamilton2017inductive, Fowlkes1983}. The binarization maintains most of the quality again, causing a 0.5--2.1\% accuracy loss.  (ii) The memory space savings are 1.72-16X. The savings are not as much as the savings in the transductive case. The reason is that most of our space savings come from the bit representation of the values of the adjacency matrices and the node features. But the sparse representation also consists of indices of the non-empty tiles. Because most non-empty tiles contain only one non-zero value, we do not save much on the indexing data structures compared to the default CSR format. For the small graphs, the indexing data structures form only a small portion of the overall representation; but for large graphs, the portion becomes much larger. But the savings are still substantial. (iii) The speedups achieved by our versions are even more significant, ranging from 2X to 125X. The reason is that as the graphs become larger, the room for time savings also grows. Another benefit worth mentioning is that as the memory space usage is reduced substantially, the new method can load the entire graph into memory and produce the inferences of all the sampled subgraphs in one run, which can bring extra time savings than processing the subgraphs one by one. 

\begin{figure*}
  \centering
  \includegraphics[trim={0cm 7.5cm 0cm 8cm},clip,width=.8\textwidth,height=.8\textheight,keepaspectratio]{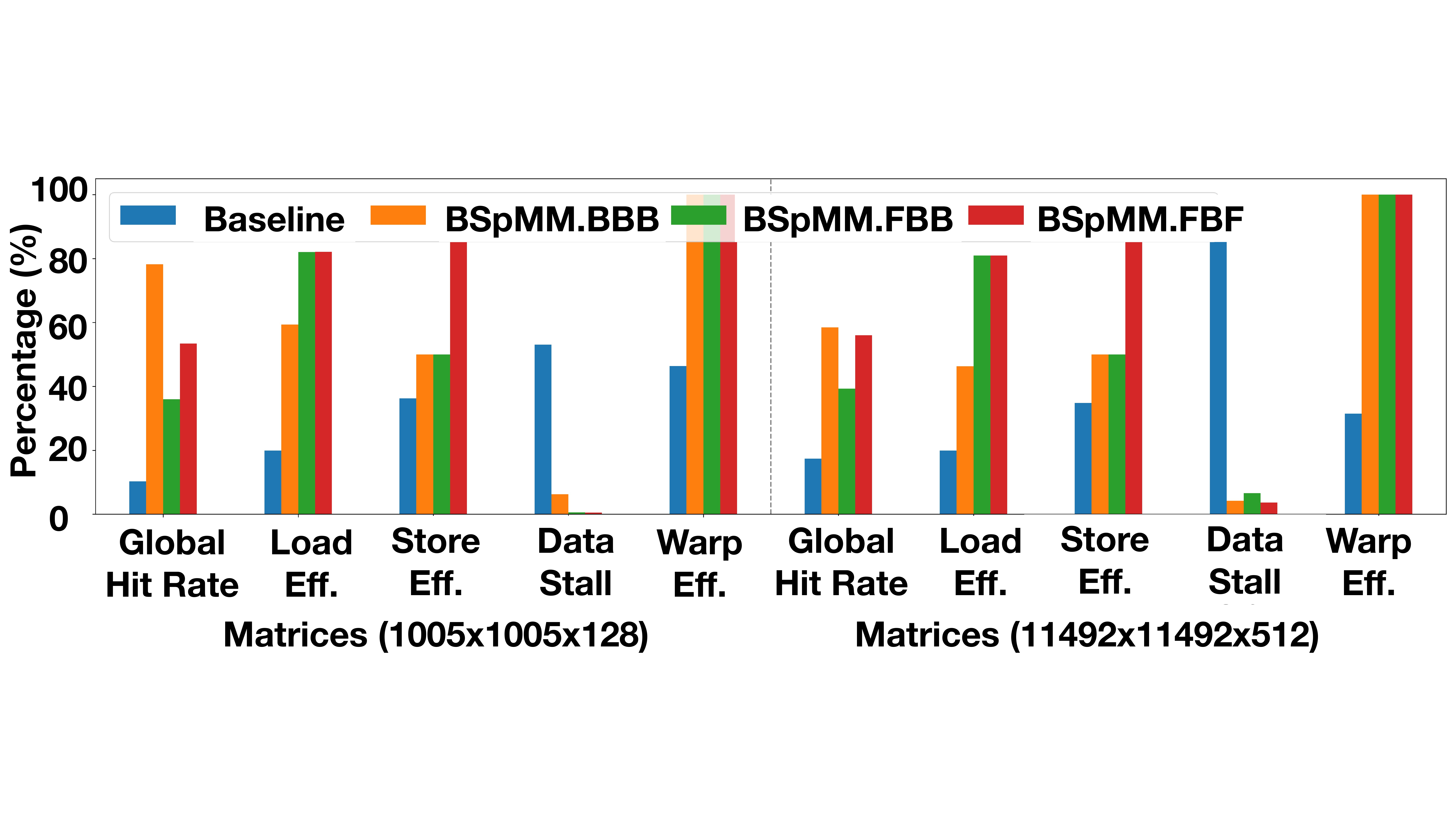}
  \caption{The profiling results of the multiplications of two pairs of sparse matrices on Nvidia GTX1080 GPU.}
  \label{eval-graph-profiling}
\end{figure*}

\begin{table*}
\centering
\caption{Evaluation results of inductive GNNs --- inductive GCN and GraphSAGE~\cite{hamilton2017inductive}. F1-micro is the metric to measure the accuracy of inductive GNNs. The {\em hidden size} is 256 for Flicker; 128 for Reddit. To avoid out-of-memory error, FP32 and Bi-GCN inferences are set to their maximal batch size on different GPUs.}
\label{tab-inductive}
\begin{tabular}{cccccccccccc}
\hline
\multirow{3}{*}{\textbf{Dataset}} & \multirow{3}{*}{\textbf{Model}}       & \multicolumn{5}{c}{\textbf{Inductive GCN}}                                                                                                                                                                                 & \multicolumn{5}{c}{\textbf{Graph SAGE}}                                                                                                                                                                                    \\ \cline{3-12} 
                                  &                                       & \multirow{2}{*}{\textbf{\begin{tabular}[c]{@{}c@{}}Peak\\ Mem (B)\end{tabular}}} & \multirow{2}{*}{\textbf{\begin{tabular}[c]{@{}c@{}}F1-\\ micro\end{tabular}}} & \multicolumn{3}{c}{\textbf{End-to-end Time (s)}}        & \multirow{2}{*}{\textbf{\begin{tabular}[c]{@{}c@{}}Peak\\ Mem (B)\end{tabular}}} & \multirow{2}{*}{\textbf{\begin{tabular}[c]{@{}c@{}}F1-\\ micro\end{tabular}}} & \multicolumn{3}{c}{\textbf{End-to-end Time (s)}}        \\ \cline{5-7} \cline{10-12} 
                                  &                                       &                                                                                  &                                                                               & \textbf{GTX1080} & \textbf{TitanV} & \textbf{RTX3060Ti} &                                                                                  &                                                                               & \textbf{GTX1080} & \textbf{TitanV} & \textbf{RTX3060Ti} \\ \hline
\multirow{5}{*}{\textbf{Flickr}}  & FP32 (S)~\cite{hamilton2017inductive} & 358.07M                                                                          & 50.9$\pm$0.3                                                                    & 0.36$\pm$0.005     & 0.33$\pm$0.007    & 0.32$\pm$0.009       & 448.37M                                                                          & 50.9$\pm$1.0                                                                    & 0.36$\pm$0.006     & 0.32$\pm$0.004    & 0.32$\pm$0.008       \\
                                  & FP32 (T)~\cite{hamilton2017inductive} & 358.07M                                                                          & 50.9$\pm$0.3                                                                    & 0.045$\pm$0.000    & 0.035$\pm$0.000   & 0.03$\pm$0.000       & 448.37M                                                                          & 50.9$\pm$1.0                                                                    & 0.045$\pm$0.001    & 0.035$\pm$0.000   & 0.034$\pm$0.000      \\
                                  & Bi-GCN~\cite{wang2021bi}              & 358.07M                                                                          & 50.2$\pm$0.4                                                                    & 0.36$\pm$0.007     & 0.34$\pm$0.010    & 0.34$\pm$0.011       & 448.37M                                                                          & 50.2$\pm$0.4                                                                    & 0.36$\pm$0.008     & 0.33$\pm$0.003    & 0.33$\pm$0.002       \\
                                  & \textbf{Ours (full)}                  & 107.07M                                                                          & 50.2$\pm$0.4                                                                    & 0.023$\pm$0.01      & 0.016$\pm$0.000    & 0.02$\pm$0.003       & 196.9M                                                                           & 50.2$\pm$0.4                                                                    & 0.038$\pm$0.007      & 0.027$\pm$0.00    & 0.03$\pm$0.006      \\
                                  & \textbf{Ours (bin)}                   & 22.38M                                                                           & 49.9$\pm$0.3                                                                    & 0.006$\pm$0.004    & 0.002$\pm$0.000   & 0.005$\pm$0.000      & 27.52M                                                                           & 50.8$\pm$0.1                                                                    & 0.008$\pm$0.008    & 0.003$\pm$0.00   & 0.006$\pm$0.001      \\ \hline
\multirow{5}{*}{\textbf{Reddit}}  & FP32 (S)                              & 1.67G                                                                            & 93.3$\pm$0.0                                                                    & 49.01$\pm$0.25     & 38.19$\pm$0.10    & 38.70$\pm$0.28       & 1.81G                                                                            & 95.3$\pm$0.1                                                                    & 49.27$\pm$0.18     & 37.95$\pm$0.35    & 38.29$\pm$0.22       \\
                                  & FP32 (T)                              & 1.67G                                                                            & 93.3$\pm$0.0                                                                    & 0.86$\pm$0.02      & 0.45$\pm$0.001    & 0.52$\pm$0.007       & 1.81G                                                                            & 95.3$\pm$0.1                                                                    & 0.92$\pm$0.001     & 0.51$\pm$0.001    & 0.52$\pm$0.002       \\
                                  & Bi-GCN                                & 1.67G                                                                            & 93.1$\pm$0.2                                                                    & 49.30$\pm$0.28     & 38.22$\pm$0.10    & 38.97$\pm$0.34       & 1.81G                                                                            & 95.3$\pm$0.1                                                                    & 49.42$\pm$0.21     & 38.05$\pm$0.56    & 39.03$\pm$0.11       \\
                                  & \textbf{Ours (full)}                  & 1.02G                                                                            & 93.1$\pm$0.2                                                                    & 0.82$\pm$0.02      & 0.38$\pm$0.08     & 0.45$\pm$0.03        & 1.17G                                                                            & 95.3$\pm$0.1                                                                    & 0.90$\pm$0.003      & 0.45$\pm$0.03     & 0.53$\pm$0.04        \\
                                  & \textbf{Ours (bin)}                   & 943.77M                                                                          & 92.8$\pm$0.1                                                                    & 0.49$\pm$0.11      & 0.15$\pm$0.007     & 0.27$\pm$0.12        & 983.77M                                                                          & 91.2$\pm$1.2                                                                    & 0.53$\pm$0.05      & 0.20$\pm$0.009     & 0.29$\pm$0.05        \\ \hline
\end{tabular}
\end{table*}

\begin{table}
\caption{Evaluation results of inductive GNNs --- GraphSAINT~\cite{graphsaintipdps19, graphsainticlr20}. The {\em hidden size} is 256 for Flicker; 128 for Reddit.}\label{tab-saint}
\resizebox{\columnwidth}{!}{
\begin{tabular}{@{}ccccccc@{}}
\toprule
\multirow{2}{*}{\textbf{Dataset}} & \multirow{2}{*}{\textbf{Model}}                     & \multirow{2}{*}{\textbf{\begin{tabular}[c]{@{}c@{}}Peak\\ Mem (B)\end{tabular}}} & \multirow{2}{*}{\textbf{\begin{tabular}[c]{@{}c@{}}F1-\\ micro\end{tabular}}} & \multicolumn{3}{c}{\textbf{End-to-end Time (s)}}        \\ \cmidrule(l){5-7} 
                                  &                                                     &                                                                                  &                                                                               & \textbf{GTX1080} & \textbf{TitanV} & \textbf{RTX3060Ti} \\ \midrule
\multirow{5}{*}{\textbf{Flickr}}  & FP32 (S)~\cite{graphsaintipdps19, graphsainticlr20} & 623.72M                                                                          & 51.1$\pm$0.1                                                                    & 0.39$\pm$0.004     & 0.36$\pm$0.003    & 0.34$\pm$0.006       \\
                                  & FP32 (T)~\cite{graphsaintipdps19, graphsainticlr20} & 623.72M                                                                          & 51.1$\pm$0.1                                                                    & 0.05$\pm$0.000     & 0.03$\pm$0.000    & 0.04$\pm$0.000       \\
                                  & Bi-GCN~\cite{wang2021bi}                            & 623.72M                                                                          & 50.8$\pm$0.2                                                                    & 0.05$\pm$0.003     & 0.04$\pm$0.000    & 0.04$\pm$0.001       \\
                                  & \textbf{Ours (full)}                                & 371.83M                                                                          & 50.8$\pm$0.2                                                                    & 0.15$\pm$0.004     & 0.07$\pm$0.00    & 0.09$\pm$0.001       \\
                                  & \textbf{Ours (bin)}                                 & 202.39M                                                                          & 49.6$\pm$0.8                                                                    & 0.023$\pm$0.007     & 0.012$\pm$0.00    & 0.016$\pm$0.001       \\ \midrule
\multirow{5}{*}{\textbf{Reddit}}  & FP32 (S)                                            & 2.04G                                                                            & 96.6$\pm$0.1                                                                    & 48.79$\pm$0.23     & 38.41$\pm$0.32    & 39.77$\pm$0.22       \\
                                  & FP32 (T)                                            & 2.04G                                                                            & 96.6$\pm$0.1                                                                    & 1.25$\pm$0.005     & 0.60$\pm$0.001    & 0.69$\pm$0.003       \\
                                  & Bi-GCN                                              & 2.04G                                                                            & 95.7$\pm$0.1                                                                    & 1.27$\pm$0.002     & 0.61$\pm$0.003    & 0.73$\pm$0.011       \\
                                  & \textbf{Ours (full)}                                & 1.39G                                                                            & 95.7$\pm$0.1                                                                    & 0.92$\pm$0.46      & 0.66$\pm$0.08     & 0.74$\pm$0.08        \\
                                  & \textbf{Ours (bin)}                                 & 1.18G                                                                            & 94.5$\pm$0.6                                                                    & 0.45$\pm$0.16      & 0.24$\pm$0.003     & 0.33$\pm$0.09        \\ \bottomrule
\end{tabular}
}
\end{table}

Additionally, Table~\ref{tab-inductive} shows a 4.1\% accuracy loss on GraphSAGE-Reddit. We attribute this to the sensitivity of the originally-trained binary GraphSAGE and the small hidden size used. While the Reddit graph is significantly larger than Flickr, it only uses as few as 128 hidden nodes. Increasing the hidden nodes to 256 and 512 results in a smaller accuracy loss ($\sim$0.67\%, $\sim$0.9\%). On Flickr, there is even an 0.6\% accuracy increase. GNN inference accuracy is unstable, especially on large graphs, which is not unique to GNNs. Studies have shown that a compressed DNN can sometimes achieve higher accuracy than the original. Our slight accuracy increase resulted from the binary aggregation approximation in the first GCN layer.

Table~\ref{tab-saint} reports the results on GraphSAINT. For this GNN, patterns of the two graphs (only one non-zero in most non-zero tiles) favor tensor kernel-based implementations. Ours (bin) still achieves consistently substantial speedups across the architecture and the datasets. It benefits from our implementation that loads only the elements corresponding to non-zero elements in the tile. The space savings are still substantial. 

\noindent\underline{\textbf{Hardware profiling:}}
Using the GPU vendor's profiling tool NSight, we find that the performance gain mainly originates from improved memory utilization through binarization. Figure~\ref{eval-graph-profiling} shows the profiling results of three of our BSpMM kernels compared to the baseline---the default SpMM kernel in Torch Sparse when computing the multiplications of two pairs of representative matrices, measured on Nvidia GTX1080 GPU. Specifically, we observe significantly enhanced global load/store efficiency (from 19\% to 80\%), and L1/Tex cache hit rate (from 10\% to 53-78\%). In addition to that, with the data request stalls reduced from 53-89\% to less than 10\%, the warp execution efficiency surges from 31-46\% to almost 100\%. 

\noindent\underline{\textbf{Discussions:}}
Regarding the overall number of operations to be executed, although BitGNN aggregates the multiply-accumulate operations of 32 elements into a single bit-dot-product operation upon one 32-bit operand, the actual reduction in operation count is less than 32X due to extra overhead on indexing. Nevertheless, since SpMM is typically memory-bound, binarization can still contribute to significant performance improvement.

\noindent\underline{\textbf{Potential impact:}}  The reduced space and time brought by BitGNN not only improves the computing efficiency on server GPUs, but also opens new opportunities for dealing with large graphs (which cannot fit into memory) and for GNN to be used on resource-constrained devices. Quantized (non-GNN) deep neural networks have already been adopted in many domains on resource-constrained embedded scenarios, including auto-driving \cite{chen2020gpu}, smart agriculture \cite{huang2021fpga}, COVID19 face-cover detection \cite{fasfous2021binarycop}, 3D object detection \cite{ma2018binary} and image processing \cite{ma2019efficient}. This is mainly due to the simplified logic \cite{geng2019lp}, low energy cost \cite{geng2020o3bnn}, low hardware requirement, and robustness \cite{galloway2017attacking}. Nevertheless, there are many scenarios the input data are non-Euclidean or can be better expressed by graphs, such as molecules \cite{wieder2020compact}, traffic flow \cite{wang2020traffic}, power-grid \cite{nauck2022predicting}, etc. In these scenarios, the binary graph neural network, as a replacement of BNN, can play a vital role in feature extraction and real-time prediction.

\section{Related Works}

\paragraph{Binary Graph Neural Network}
Binary GNNs have recently gained attention as a promising approach for applying quantization to graph learning workloads in the ML community. Despite several works proposing GNN binarization techniques and/or binary GNN models~\cite{wang2021bi, bahri2021binary, wang2021binarized, jing2021meta}, none of them have demonstrated the full performance advantages that can be achieved through binarization, which is where BitGNN comes in. Here, we summarize some of the existing binary GNN studies.

Binary Graph Convolutional Network (Bi-GCN)~\cite{wang2021bi} proposes a novel binary gradient approximation-based back-propagation technique for effectively training binary GCNs. It binarizes both the network parameters and node features with minimal accuracy loss. Bi-GCN has been applied to both transductive-learning GNNs (e.g., GCN~\cite{kipf2017semi}) and inductive-learning GNNs (e.g., inductive-GCN, GraphSAGE~\cite{hamilton2017inductive}, and GraphSAINT~\cite{graphsaintipdps19, graphsainticlr20}). Alternatively, Bahri et al.~\cite{bahri2021binary} use knowledge distillation and multi-stage training to binarize the EdgeConv in Dynamic Graph CNNs (DGCNN)~\cite{wang2019dynamic}. Wang et al.~\cite{wang2021binarized} propose a new binarized graph embedding method, named BGN, to binarize the parameters and pre-activations of Adaptive Sampling (AS-GCN)~\cite{huang2018adaptive} and Graph Attention Network (GAT)~\cite{vel2018graph}. In the back-propagation phase, they use two popular unbiased gradient estimators --- the straight-through estimator and REINFORCE estimator~\cite{williams1992simple} --- for binary approximation. Aside from XNOR and population count for +1 and -1 bit calculation, they involve masked summation and balance functions to improve the binarization process. Meta-Aggregator~\cite{jing2021meta} introduces two aggregators --- the Greedy Gumbel Aggregator (GNA) and Adaptable Hybrid Aggregator (ANA) --- aiming at enhancing the binary training accuracy during the aggregation phase.

While these binarization techniques for GNNs provide a stepping stone in exploring the potential of binary GNNs, they only "logically" binarize GNN at the algorithm level, and their implementations still use \texttt{sign()} and \texttt{mm()} in PyTorch, which essentially adopt full-precision tensors for storage and full-precision operations for computation at the lower level. As a result, they fail to showcase the performance potential of binary GNNs.

\paragraph{Bit-Manipulation on GPUs}
Previous research has extensively explored the potential of leveraging bit operations on modern GPUs across a wide range of application scenarios. These include image classification using CNNs~\cite{courbariaux2016binarized, pedersoli2018espresso, khan2018binarized, li2019bstc, li2020accelerating}, depth-sensing with stereo vision~\cite{chen2020gpu, meng2021gpu}, fraud detection~\cite{chang2021fraud}, and graph algorithms~\cite{chen2022bitgraphblas}.

Courbariaux et al.~\cite{courbariaux2016binarized} first introduce the concept of using XOR and population-count operations for small MLPs on GPUs. Espresso\\~\cite{pedersoli2018espresso} builds a binary CNN library with C and CUDA backends, providing bit-packing and bit matrix multiplication functions. Khan et al.~\cite{khan2018binarized} extend the previous GEMM-based CNNs for real-time vehicle classification on desktop and embedded GPUs. BSTC~\cite{li2019bstc} presents bit-block-based kernel designs for BMM and BConv, scaling the significant performance gains to deeper binary AlexNet, VGGNet, and ResNet. StereoBit~\cite{chen2020gpu} and FastFusion~\cite{meng2021gpu} develop BNN-based stereo matching networks with optimized implementations to reduce GPU on-chip memory footprint. They construct the network with binary feature descriptors for image pairs and weights, replacing expensive full-precision arithmetic operations with XOR and population count operations. Ye et al.~\cite{chang2021fraud} leverage the GPU intrinsics \_\_ballot\_sync(), \_\_match\_any\_sync(), and \_\_popc() to construct a Label Propagation (LP) algorithm for the fraud detection pipeline. Feng et al.~\cite{feng2021apnn} decompose quantized neural network matrix multiplications into batches of BMM operations and relied on the Ampere tensor cores for acceleration. Finally, Bit-GraphBLAS~\cite{chen2022bitgraphblas} proposes a sparse bit storage format and binary implementation of GraphBLAS operators to accelerate iteration-based graph algorithms when the graph edges are homogeneous and can be expressed as binary graphs.
 
These prior studies demonstrate the performance benefits of utilizing bit manipulations and operations on GPUs. However, they are not comprehensive enough to address the specific challenges that arise when dealing with binary GNNs, such as the need for operator restructuring, appropriate tensor representation, and inconsistent value multiplications.

\section{Conclusion}
BitGNN is the first work to comprehensively address the design and challenges of implementing binary GNNs using bit tensors and manipulations. Our proposed programming abstractions, tensor representation, and kernel design techniques maximize the performance of binary GNN models end-to-end. The optimizations and implementations provide a remarkable 8-22X acceleration in the end-to-end model performance compared to prior binary GNN implementations.

\section*{Acknowledgement}
This material is based upon work supported by the U.S. Department of Energy (Office of Advanced Scientific Computing Research)  
under Award Numbers DE-EE0009357 and 78284, as well as work supported by the National Science Foundation (NSF) under Grant No. CNS-2107068. Any opinions, findings, and conclusions or recommendations expressed in this material are those of the authors and do not necessarily reflect the views of NSF or DOE.

\bibliographystyle{unsrt}
\bibliography{main}

\begin{thebibliography}{10}

\bibitem{fan2019graph}
Wenqi Fan, Yao Ma, Qing Li, Yuan He, Eric Zhao, Jiliang Tang, and Dawei Yin.
\newblock Graph neural networks for social recommendation.
\newblock In {\em The world wide web conference}, pages 417--426, 2019.

\bibitem{reau2023deeprank}
Manon R{\'e}au, Nicolas Renaud, Li~C Xue, and Alexandre~MJJ Bonvin.
\newblock Deeprank-gnn: a graph neural network framework to learn patterns in
  protein--protein interfaces.
\newblock {\em Bioinformatics}, 39(1):btac759, 2023.

\bibitem{gilmer2017neural}
Justin Gilmer, Samuel~S Schoenholz, Patrick~F Riley, Oriol Vinyals, and
  George~E Dahl.
\newblock Neural message passing for quantum chemistry.
\newblock In {\em International conference on machine learning}, pages
  1263--1272. PMLR, 2017.

\bibitem{helal2022extreme}
Hatem Helal, Jesun Firoz, Jenna Bilbrey, Mario~Michael Krell, Tom Murray, Ang
  Li, Sotiris Xantheas, and Sutanay Choudhury.
\newblock Extreme acceleration of graph neural network-based prediction models
  for quantum chemistry.
\newblock {\em arXiv preprint arXiv:2211.13853}, 2022.

\bibitem{wang2019dynamic}
Yue Wang, Yongbin Sun, Ziwei Liu, Sanjay~E. Sarma, Michael~M. Bronstein, and
  Justin~M. Solomon.
\newblock Dynamic graph cnn for learning on point clouds.
\newblock {\em ACM Trans. Graph.}, 38(5), oct 2019.

\bibitem{zhang2021pointx}
Jie-Fang Zhang and Zhengya Zhang.
\newblock Point-x: A spatial-locality-aware architecture for energy-efficient
  graph-based point-cloud deep learning.
\newblock In {\em MICRO-54: 54th Annual IEEE/ACM International Symposium on
  Microarchitecture}, page 1078–1090, New York, NY, USA, 2021. Association
  for Computing Machinery.

\bibitem{kipf2017semi}
Thomas~N. Kipf and Max Welling.
\newblock Semi-supervised classification with graph convolutional networks.
\newblock In {\em International Conference on Learning Representations (ICLR)},
  2017.

\bibitem{yao2019graph}
Liang Yao, Chengsheng Mao, and Yuan Luo.
\newblock Graph convolutional networks for text classification.
\newblock In {\em Proceedings of the AAAI conference on artificial
  intelligence}, volume~33, pages 7370--7377, 2019.

\bibitem{ye2019web}
Hongfan Ye, Buqing Cao, Junjie Chen, Jianxun Liu, Yiping Wen, and Jinjun Chen.
\newblock A web services classification method based on gcn.
\newblock In {\em 2019 IEEE Intl Conf on Parallel \& Distributed Processing
  with Applications, Big Data \& Cloud Computing, Sustainable Computing \&
  Communications, Social Computing \& Networking
  (ISPA/BDCloud/SocialCom/SustainCom)}, pages 1107--1114. IEEE, 2019.

\bibitem{liang2020deep}
Jiali Liang, Yufan Deng, and Dan Zeng.
\newblock A deep neural network combined cnn and gcn for remote sensing scene
  classification.
\newblock {\em IEEE Journal of Selected Topics in Applied Earth Observations
  and Remote Sensing}, 13:4325--4338, 2020.

\bibitem{wan2020hyperspectral}
Sheng Wan, Chen Gong, Ping Zhong, Shirui Pan, Guangyu Li, and Jian Yang.
\newblock Hyperspectral image classification with context-aware dynamic graph
  convolutional network.
\newblock {\em IEEE Transactions on Geoscience and Remote Sensing},
  59(1):597--612, 2020.

\bibitem{jiang2020hi}
Hao Jiang, Peng Cao, MingYi Xu, Jinzhu Yang, and Osmar Zaiane.
\newblock Hi-gcn: A hierarchical graph convolution network for graph embedding
  learning of brain network and brain disorders prediction.
\newblock {\em Computers in Biology and Medicine}, 127:104096, 2020.

\bibitem{hubara2016binarized}
Itay Hubara, Matthieu Courbariaux, Daniel Soudry, Ran El-Yaniv, and Yoshua
  Bengio.
\newblock Binarized neural networks.
\newblock {\em Advances in neural information processing systems}, 29, 2016.

\bibitem{courbariaux2016binarized}
Matthieu Courbariaux, Itay Hubara, Daniel Soudry, Ran El-Yaniv, and Yoshua
  Bengio.
\newblock Binarized neural networks: Training deep neural networks with weights
  and activations constrained to +1 or -1.
\newblock {\em arXiv preprint arXiv:1602.02830}, 2016.

\bibitem{rastegari2016xnor}
Mohammad Rastegari, Vicente Ordonez, Joseph Redmon, and Ali Farhadi.
\newblock Xnor-net: Imagenet classification using binary convolutional neural
  networks.
\newblock In {\em European conference on computer vision}, pages 525--542.
  Springer, 2016.

\bibitem{bulat2019xnor}
Adrian Bulat and Georgios Tzimiropoulos.
\newblock Xnor-net++: Improved binary neural networks.
\newblock In {\em Proceedings of the British Machine Vision Conference (BMVC)},
  2019.

\bibitem{zhang2022pokebnn}
Yichi Zhang, Zhiru Zhang, and Lukasz Lew.
\newblock Pokebnn: A binary pursuit of lightweight accuracy.
\newblock In {\em Proceedings of the IEEE/CVF Conference on Computer Vision and
  Pattern Recognition}, pages 12475--12485, 2022.

\bibitem{wang2021bi}
Junfu Wang, Yunhong Wang, Zhen Yang, Liang Yang, and Yuanfang Guo.
\newblock Bi-gcn: Binary graph convolutional network.
\newblock In {\em Proceedings of the IEEE/CVF Conference on Computer Vision and
  Pattern Recognition}, pages 1561--1570, 2021.

\bibitem{bahri2021binary}
Mehdi Bahri, Ga{\'e}tan Bahl, and Stefanos Zafeiriou.
\newblock Binary graph neural networks.
\newblock In {\em Proceedings of the IEEE/CVF Conference on Computer Vision and
  Pattern Recognition}, pages 9492--9501, 2021.

\bibitem{wang2021binarized}
Hanchen Wang, Defu Lian, Ying Zhang, Lu~Qin, Xiangjian He, Yiguang Lin, and
  Xuemin Lin.
\newblock Binarized graph neural network.
\newblock {\em World Wide Web}, 24(3):825--848, 2021.

\bibitem{jing2021meta}
Yongcheng Jing, Yiding Yang, Xinchao Wang, Mingli Song, and Dacheng Tao.
\newblock Meta-aggregator: Learning to aggregate for 1-bit graph neural
  networks.
\newblock In {\em Proceedings of the IEEE/CVF International Conference on
  Computer Vision}, pages 5301--5310, 2021.

\bibitem{fey2019fast}
Matthias Fey and Jan~E. Lenssen.
\newblock Fast graph representation learning with {PyTorch Geometric}.
\newblock In {\em ICLR Workshop on Representation Learning on Graphs and
  Manifolds}, 2019.

\bibitem{hamilton2017inductive}
William~L Hamilton, Rex Ying, and Jure Leskovec.
\newblock Inductive representation learning on large graphs.
\newblock In {\em Proceedings of the 31st International Conference on Neural
  Information Processing Systems}, pages 1025--1035, 2017.

\bibitem{morris2019weisfeiler}
Christopher Morris, Martin Ritzert, Matthias Fey, William~L Hamilton, Jan~Eric
  Lenssen, Gaurav Rattan, and Martin Grohe.
\newblock Weisfeiler and leman go neural: Higher-order graph neural networks.
\newblock In {\em Proceedings of the AAAI Conference on Artificial
  Intelligence}, volume~33, pages 4602--4609, 2019.

\bibitem{graphsaintipdps19}
Hanqing Zeng, Hongkuan Zhou, Ajitesh Srivastava, Rajgopal Kannan, and Viktor
  Prasanna.
\newblock Accurate, efficient and scalable graph embedding.
\newblock In {\em 2019 IEEE International Parallel and Distributed Processing
  Symposium (IPDPS)}, May 2019.

\bibitem{graphsainticlr20}
Hanqing Zeng, Hongkuan Zhou, Ajitesh Srivastava, Rajgopal Kannan, and Viktor
  Prasanna.
\newblock Graphsaint: Graph sampling based inductive learning method.
\newblock In {\em International Conference on Learning Representations}, 2020.

\bibitem{li2018warp}
Ang Li, Weifeng Liu, Linnan Wang, Kevin Barker, and Shuaiwen~Leon Song.
\newblock Warp-consolidation: A novel execution model for gpus.
\newblock In {\em Proceedings of the 2018 International Conference on
  Supercomputing}, ICS '18, page 53–64, New York, NY, USA, 2018. Association
  for Computing Machinery.

\bibitem{li2019bstc}
Ang Li, Tong Geng, Tianqi Wang, Martin Herbordt, Shuaiwen~Leon Song, and Kevin
  Barker.
\newblock Bstc: A novel binarized-soft-tensor-core design for accelerating
  bit-based approximated neural nets.
\newblock In {\em Proceedings of the International Conference for High
  Performance Computing, Networking, Storage and Analysis}, SC '19, New York,
  NY, USA, 2019. Association for Computing Machinery.

\bibitem{amd2022hip}
AMD.
\newblock Hip programming guide:
  https://github.com/radeonopencompute/rocm/\\blob/rocm-4.5.2/amd\_hip\_programming\_guide.pdf,
  2022.

\bibitem{chen2022bitgraphblas}
Jou-An Chen, Hsin-Hsuan Sung, Xipeng Shen, Nathan Tallent, Kevin Barker, and
  Ang Li.
\newblock Bit-graphblas: Bit-level optimizations of matrix-centric graph
  processing on gpu.
\newblock In {\em 2022 IEEE International Parallel and Distributed Processing
  Symposium (IPDPS)}, pages 515--525, 2022.

\bibitem{li2020accelerating}
Ang Li and Simon Su.
\newblock Accelerating binarized neural networks via bit-tensor-cores in turing
  gpus.
\newblock {\em IEEE Transactions on Parallel and Distributed Systems},
  32(7):1878--1891, 2020.

\bibitem{ben2016fast}
Eli Ben-Sasson, Matan Hamilis, Mark Silberstein, and Eran Tromer.
\newblock Fast multiplication in binary fields on gpus via register cache.
\newblock In {\em Proceedings of the 2016 International Conference on
  Supercomputing}, pages 1--12, 2016.

\bibitem{huang2020gespmm}
Guyue Huang, Guohao Dai, Yu~Wang, and Huazhong Yang.
\newblock Ge-spmm: General-purpose sparse matrix-matrix multiplication on gpus
  for graph neural networks.
\newblock In {\em Proceedings of the International Conference for High
  Performance Computing, Networking, Storage and Analysis}, SC '20. IEEE Press,
  2020.

\bibitem{rahman2020fusedmm}
Md~Rahman, Majedul~Haque Sujon, Ariful Azad, et~al.
\newblock Fusedmm: A unified sddmm-spmm kernel for graph embedding and graph
  neural networks.
\newblock In {\em 35th Proceedings of IEEE IPDPS}, 2021.

\bibitem{bitgnnapendix}
Technical report -- bitgnn: Unleashing the performance potential of binary
  graph neural networks on gpus: https://tinyurl.com/yuf87cax, 2023.

\bibitem{cudaprogrammingguide2022}
NVIDIA.
\newblock Cuda programming guide, 2022.

\bibitem{zhao2018}
Yue Zhao, Jiajia Li, Chunhua Liao, and Xipeng Shen.
\newblock Bridging the gap between deep learning and sparse matrix format
  selection.
\newblock In {\em Proceedings of the 23rd ACM SIGPLAN Symposium on Principles
  and Practice of Parallel Programming}, 2018.

\bibitem{yang2016revisiting}
Zhilin Yang, William Cohen, and Ruslan Salakhudinov.
\newblock Revisiting semi-supervised learning with graph embeddings.
\newblock In {\em International conference on machine learning}, pages 40--48.
  PMLR, 2016.

\bibitem{hu2020open}
Weihua Hu, Matthias Fey, Marinka Zitnik, Yuxiao Dong, Hongyu Ren, Bowen Liu,
  Michele Catasta, and Jure Leskovec.
\newblock Open graph benchmark: Datasets for machine learning on graphs.
\newblock {\em Advances in neural information processing systems},
  33:22118--22133, 2020.

\bibitem{shchur2018pitfalls}
Oleksandr Shchur, Maximilian Mumme, Aleksandar Bojchevski, and Stephan
  G{\"u}nnemann.
\newblock Pitfalls of graph neural network evaluation.
\newblock {\em arXiv preprint arXiv:1811.05868}, 2018.

\bibitem{zitnik2017predicting}
Marinka Zitnik and Jure Leskovec.
\newblock Predicting multicellular function through multi-layer tissue
  networks.
\newblock {\em Bioinformatics}, 33(14):i190--i198, 2017.

\bibitem{Fowlkes1983}
R.~J. Fowlkes and D.~G. Mallows.
\newblock A method for comparing two hierarchical clusterings.
\newblock {\em Journal of the American Statistical Association},
  78(383):553--569, 1983.

\bibitem{chen2020gpu}
Gang Chen, Haitao Meng, Yucheng Liang, and Kai Huang.
\newblock Gpu-accelerated real-time stereo estimation with binary neural
  network.
\newblock {\em IEEE Transactions on Parallel and Distributed Systems},
  31(12):2896--2907, 2020.

\bibitem{huang2021fpga}
Chun-Hsian Huang.
\newblock An fpga-based hardware/software design using binarized neural
  networks for agricultural applications: A case study.
\newblock {\em IEEE Access}, 9:26523--26531, 2021.

\bibitem{fasfous2021binarycop}
Nael Fasfous, Manoj-Rohit Vemparala, Alexander Frickenstein, Lukas
  Frickenstein, Mohamed Badawy, and Walter Stechele.
\newblock Binarycop: Binary neural network-based covid-19 face-mask wear and
  positioning predictor on edge devices.
\newblock In {\em 2021 IEEE International Parallel and Distributed Processing
  Symposium Workshops (IPDPSW)}, pages 108--115. IEEE, 2021.

\bibitem{ma2018binary}
Chao Ma, Yulan Guo, Yinjie Lei, and Wei An.
\newblock Binary volumetric convolutional neural networks for 3-d object
  recognition.
\newblock {\em IEEE Transactions on Instrumentation and Measurement},
  68(1):38--48, 2018.

\bibitem{ma2019efficient}
Yinglan Ma, Hongyu Xiong, Zhe Hu, and Lizhuang Ma.
\newblock Efficient super resolution using binarized neural network.
\newblock In {\em Proceedings of the IEEE/CVF Conference on Computer Vision and
  Pattern Recognition Workshops}, pages 0--0, 2019.

\bibitem{geng2019lp}
Tong Geng, Tianqi Wang, Chunshu Wu, Chen Yang, Shuaiwen~Leon Song, Ang Li, and
  Martin Herbordt.
\newblock Lp-bnn: Ultra-low-latency bnn inference with layer parallelism.
\newblock In {\em 2019 IEEE 30th International Conference on
  Application-specific Systems, Architectures and Processors (ASAP)}, volume
  2160, pages 9--16. IEEE, 2019.

\bibitem{geng2020o3bnn}
Tong Geng, Ang Li, Tianqi Wang, Chunshu Wu, Yanfei Li, Runbin Shi, Wei Wu, and
  Martin Herbordt.
\newblock O3bnn-r: An out-of-order architecture for high-performance and
  regularized bnn inference.
\newblock {\em IEEE Transactions on parallel and distributed systems},
  32(1):199--213, 2020.

\bibitem{galloway2017attacking}
Angus Galloway, Graham~W Taylor, and Medhat Moussa.
\newblock Attacking binarized neural networks.
\newblock {\em arXiv preprint arXiv:1711.00449}, 2017.

\bibitem{wieder2020compact}
Oliver Wieder, Stefan Kohlbacher, M{\'e}laine Kuenemann, Arthur Garon, Pierre
  Ducrot, Thomas Seidel, and Thierry Langer.
\newblock A compact review of molecular property prediction with graph neural
  networks.
\newblock {\em Drug Discovery Today: Technologies}, 37:1--12, 2020.

\bibitem{wang2020traffic}
Xiaoyang Wang, Yao Ma, Yiqi Wang, Wei Jin, Xin Wang, Jiliang Tang, Caiyan Jia,
  and Jian Yu.
\newblock Traffic flow prediction via spatial temporal graph neural network.
\newblock In {\em Proceedings of the web conference 2020}, pages 1082--1092,
  2020.

\bibitem{nauck2022predicting}
Christian Nauck, Michael Lindner, Konstantin Sch{\"u}rholt, Haoming Zhang, Paul
  Schultz, J{\"u}rgen Kurths, Ingrid Isenhardt, and Frank Hellmann.
\newblock Predicting basin stability of power grids using graph neural
  networks.
\newblock {\em New Journal of Physics}, 24(4):043041, 2022.

\bibitem{huang2018adaptive}
Wenbing Huang, Tong Zhang, Yu~Rong, and Junzhou Huang.
\newblock Adaptive sampling towards fast graph representation learning.
\newblock In S.~Bengio, H.~Wallach, H.~Larochelle, K.~Grauman, N.~Cesa-Bianchi,
  and R.~Garnett, editors, {\em Advances in Neural Information Processing
  Systems}, volume~31. Curran Associates, Inc., 2018.

\bibitem{vel2018graph}
Petar Veličković, Guillem Cucurull, Arantxa Casanova, Adriana Romero, Pietro
  Liò, and Yoshua Bengio.
\newblock Graph attention networks.
\newblock In {\em International Conference on Learning Representations}, 2018.

\bibitem{williams1992simple}
Ronald~J Williams.
\newblock Simple statistical gradient-following algorithms for connectionist
  reinforcement learning.
\newblock {\em Machine learning}, 8(3):229--256, 1992.

\bibitem{pedersoli2018espresso}
Fabrizio Pedersoli, George Tzanetakis, and Andrea Tagliasacchi.
\newblock Espresso: Efficient forward propagation for binary deep neural
  networks.
\newblock In {\em International Conference on Learning Representations}, 2018.

\bibitem{khan2018binarized}
Mir Khan, Heikki Huttunen, and Jani Boutellier.
\newblock Binarized convolutional neural networks for efficient inference on
  gpus.
\newblock In {\em 2018 26th European Signal Processing Conference (EUSIPCO)},
  pages 682--686. IEEE, 2018.

\bibitem{meng2021gpu}
Haitao Meng, Chonghao Zhong, Jianfeng Gu, and Gang Chen.
\newblock A gpu-accelerated deep stereo-lidar fusion for real-time
  high-precision dense depth sensing.
\newblock In {\em 2021 Design, Automation \& Test in Europe Conference \&
  Exhibition (DATE)}, pages 523--528. IEEE, 2021.

\bibitem{chang2021fraud}
Chang Ye, Yuchen Li, Bingsheng He, Zhao Li, and Jianling Sun.
\newblock {\em GPU-Accelerated Graph Label Propagation for Real-Time Fraud
  Detection}, page 2348–2356.
\newblock Association for Computing Machinery, New York, NY, USA, 2021.

\bibitem{feng2021apnn}
Boyuan Feng, Yuke Wang, Tong Geng, Ang Li, and Yufei Ding.
\newblock Apnn-tc: Accelerating arbitrary precision neural networks on ampere
  gpu tensor cores.
\newblock In {\em Proceedings of the international conference for high
  performance computing, networking, storage and analysis}, pages 1--13, 2021.

\end{thebibliography}
\end{document}